\documentclass[twocolumn,superscriptaddress,showpacs,aps,floatfix]{revtex4-1}

\usepackage[english]{babel}

\usepackage{soul}
\usepackage{orcidlink}
\usepackage{comment}

\usepackage{amsmath}
\usepackage{bm}
\usepackage{siunitx}
\usepackage{nicefrac}
\usepackage{braket}

\usepackage{graphicx}
\usepackage{xcolor}
\usepackage{xspace}
\usepackage{float}

\DeclareMathAlphabet\mathbfcal{OMS}{cmsy}{b}{n}

\usepackage{hyperref}

\newcommand{\cnrspin}{\affiliation{Consiglio Nazionale delle Ricerche CNR-SPIN, c/o Universit\'{a} degli Studi "G. D’Annunzio", 66100 Chieti, Italy}}
\newcommand{\bicocca}{\affiliation{Department of Materials Science, University of Milan-Bicocca, Via Roberto Cozzi 55, 20125 Milan, Italy}}
\newcommand{\vinca}{\affiliation{Vin\v{c}a Institute of Nuclear Sciences - 
National Institute of the Republic of Serbia, University of Belgrade, P. O. Box 522, RS-11001 Belgrade, Serbia}}
\newcommand{\cnrrome}{\affiliation{Consiglio Nazionale delle Ricerche CNR-SPIN, Area della Ricerca di Tor Vergata, Via del Fosso del Cavaliere, 100, I-00133 Rome, Italy}}
\newcommand{\cfm}{\affiliation{Centro de F\'{i}sica de Materiales (CSIC-UPV/EHU), 20018, Donostia-San Sebasti\'{a}n, Spain}}
\newcommand{\iker}{\affiliation{IKERBASQUE, Basque Foundation for Science, 48009 Bilbao, Spain}}
\newcommand{\dipc}{\affiliation{Donostia International Physics Center (DIPC), 20018
Donostia-San Sebasti\'{a}n, Spain}}
\newcommand{\unive}{\affiliation{Department of Molecular Sciences and Nanosystems, Ca’ Foscari University of Venice, via Torino 155, 30170, Venice-Mestre, Italy}}

\begin{document}
\title{Giant non-reciprocal band structure effect in a multiferroic material}

\author{Srdjan Stavri\'c\orcidlink{0000-0003-2097-0955} } \vinca \cnrspin 
\author{Giuseppe Cuono} \cnrspin
\author{Baishun Yang} \cnrspin
\author{{\'A}lvaro R. Puente-Uriona}\cfm
\author{Julen Iba\~{n}ez-Azpiroz}\cfm \iker \dipc
\author{Paolo Barone} \cnrrome
\author{Andrea Droghetti} 
\email[]{andrea.droghetti@unive.it}  \unive 
\author{Silvia Picozzi} \bicocca \cnrspin

\date{\today}

\begin{abstract}
Multiferroic materials, characterized by the coexistence of ferroelectricity and ferromagnetism, 
may unveil band structures suggestive of complex phenomena and new functionalities. In this Letter, we analyze the band structure of EuO in its multiferroic phase.
Using density functional theory calculations and detailed symmetry analysis, we reveal a previously overlooked non-reciprocal band structure effect, where the electronic energy bands exhibit asymmetry along opposite directions with respect to the special points in the Brillouin zone. This effect, which is enabled by spin-orbit coupling, is giant for the top valence Eu $4f$ bands, and can be switched by external electric or magnetic fields. Furthermore, this results in an enhanced bulk photovoltaic effect. Specifically, our predictions indicate the emergence of a large injection current response to linearly polarized light, resulting in a photoconductivity value several orders of magnitude higher than that reported in any other oxide material. Ultimately, this non-reciprocal band structure effect and the associated large bulk photovoltaic response may be general phenomena emerging not just in EuO but also in other multiferroics or magnetoelectrics, potentially providing  new cross-functionalities.

\end{abstract}

\maketitle

{\it Introduction.} Multiferroic materials exhibit coupled ferroelectric and magnetic degrees of freedom \cite{Spaldin2005,Spaldin2019}, enabling cross control of ferroic order parameters \cite{Zhao2006,Chu2008,He2011, Heron2014,Manipatruni2019,Fert2024,Kimura2003,Lee2013}.
Additionally, the simultaneous breaking of time-reversal and spatial inversion symmetries leads to nonreciprocal responses to external stimuli, introducing directional dependencies, and thus enriching their cradle of functionalities \cite{Tokura2018}.
While many mechanisms underlying the coexistence of the different ferroic orders have been identified \cite{Hill2000,Cheong2007,vdBrink2008,Tokura2014, Barone2015, Fiebig2016}, aspects related to their band structures remain comparatively less investigated. Heuristically, energy bands, $\epsilon_{n\sigma}(\bm k)$, in semiconductors lacking both time-reversal and spatial inversion symmetries would break the relation $\epsilon_{n\sigma}(\bm k)=\epsilon_{n{\pm\sigma}}(-\bm k)$, introducing asymmetries in group velocities upon reversing the wave-vector, $\bm k$ \cite{Winkler2023}. 
This can lead to light-induced dc currents from velocity differences of excited carriers, contributing to the bulk photovoltaic effect (BPE) \cite{Baltz1981,Sipe2000,Dai2023}. In ferroelectrics, the non-volatile electrical control of the BPE via spontaneous polarization has been proposed for switchable optoelectronic devices \cite{Choi2009}. In multiferroics, this functionality could be further enhanced through cross-coupling effects, for the understanding of which a band structure analysis becomes crucial.

\begin{figure*}[ht!]
\centering
\includegraphics[width=0.8\textwidth]{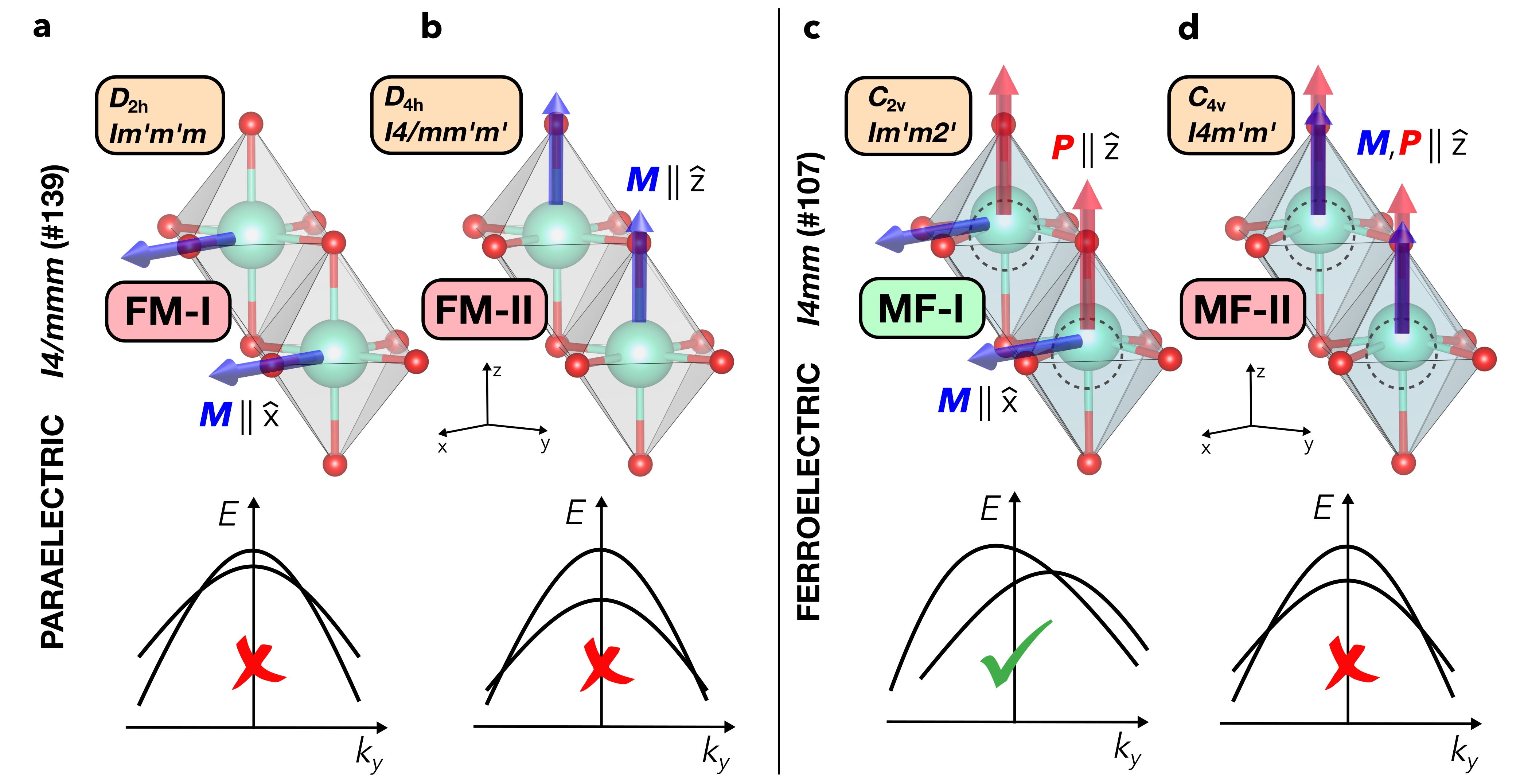}
\caption{
Schematic crystal and band structures for the FM and MF systems with $\bm{M}$ $\parallel$ or $\perp c$.
}
\label{fig:para_vs_ferro} 
\end{figure*}

In this Letter, we choose EuO in its  multiferroic (MF) phase as a paradigmatic system to analyze its band-structure as a function of magnetization, ${\bm M}$, and electrical polarization, ${\bm P}$, whose directions are entangled by spin-orbit coupling (SOC). Through density functional theory (DFT) calculations, we find that some bands, $\epsilon_n({\bm k})$, become asymmetric upon reversing $\bm k$, a phenomenon we refer to as a non-reciprocal band effect (NRBE). Magnetic and spatial symmetry operations other than inversion symmetry introduce further constraints on the band structure, shaping its dependence on the relative $\bm M$ and $\bm P$ directions. For example, when ${\bm M}$ and ${\bm P}$ are perpendicular to each other (${\bm P} \perp {\bm M}$), we obtain $\epsilon_n({\bm k})\neq \epsilon_n(-{\bm k})$ for ${\bm k}\parallel {\bm P}\times {\bm M}$ around the $\Gamma$ point in the Brillouin zone (BZ), as illustrated in Fig. \ref{fig:para_vs_ferro}. Such NRBE is giant for the valence bands, where $\vert\epsilon_n({\bm k})- \epsilon_n(-{\bm k})\vert$ exceeds $300$ meV. Furthermore, it can be controlled by external electric or magnetic fields, changing the relative orientation of ${\bm M}$ and ${\bm P}$. 
Finally, we calculate the intrinsic non-linear photoconductivity,  
linking it to the band-structure and showing that the contribution from the NRBE dominates the BPE within the visible light energy range.

{\it EuO crystal and electronic structure.} EuO is a ferromagnetic (FM) semiconductor with a rocksalt (RS)-cubic crystal and a Curie temperature of $69 \, {\mathrm K}$ \cite{Matthias1961,Ahn1968,Mauger1986}. Its ferromagnetism is due to the parallel alignment of localized magnetic moments ($=7$ $\mu_B$), resulting from the half-filled $4f$ shell of the Eu$^{2+}$ ions \cite{Passell1976,Miyazaki2009}. Due to a tiny cubic magnetocrystalline anisotropy (MCA) energy of only $0.2$ $\mu$eV/f.u. \cite{Miyata1967}, EuO is a nearly ideal material realization of the classical (isotropic) Heisenberg model \cite{Als-Nielsen1971}. 
The top valence bands are occupied majority-spin Eu $4f$ states, while the conduction bands are predominantly Eu $5d$ states, exchange-split by $\sim0.6$ eV \cite{Steeneken2002,Ghosh2004,Ingle2008,An2011},
which can act as a spin-filter in spintronic devices\cite{Santos2004,Schmehl2007,Santos2008,Jutong2012}.

EuO in its native RS structure is not a MF material but it can become one through strain engineering.
Depending on the substrate choice \cite{Iwata2000,Lettieri2003,Sutarto2009,Swartz2010}, EuO thin films either retain the RS lattice \cite{Sutarto2009} or are subjected to biaxial strain \cite{Melville2013}. 
In the latter case, first-principles calculations \cite{Bousquet2010} predicted that tensile (compressive) biaxial strain exceeding $+4.2\%$ ($-3.3\%$) induce a polar phonon instability, triggering a displacive phase transition to a low-symmetry structure with in-plane (out-of-plane) spontaneous electrical polarization. This was experimentally confirmed in (EuO)$_x$/(BaO)$_y$ superlattices, where EuO experienced a tensile strain up to $+6.4\%$ \cite{Goian2020}.

Without loss of generality, we focus on EuO under compressive biaxial strain. We perform non-collinear DFT calculations using the Vienna {\it Ab-initio} Simulation Package ({\sc VASP})~\cite{Kresse1993}, with PBE GGA exchange-correlation functional \cite{Perdew1996} and SOC included. As GGA incorrectly predicts EuO as metallic, a $U$ correction is applied to the Eu $4f$ orbitals via the approach of Ref. \cite{Dudarev1998}, with $U_{\mathrm{eff}}=U-J=7.4$ eV. Additional computational details are in section S1 of the Supporting Information (SI). 

We start with the native RS structure and apply a $-5 \%$ biaxial compressive strain in the $ab$ plane. This realizes \textit{the paraelectric} (PE) structure with $I4/mmm$ space group (No. 139, crystallographic point group $D_{4h}$). Since this strain exceeds the threshold for the displacive phase transition \cite{Bousquet2010}, further ionic relaxation ends in a low-symmetry structure with the $I4mm$ space group (No. 107, crystallographic point group $C_{4v}$). This is $5.7\, \mathrm{meV}/\mathrm{f.u.}$ lower in energy than the initial PE structure and displays a separation of the center of mass of the positive (Eu$^{2+}$) and negative (O$^{2-}$) ions, resulting in an electrical polarization along the $c$-axis, ${\bm P} \parallel c$.
Given this polarization, we refer to the $I4mm$ structure as \textit{the ferroelecric} (FE) structure.

Including the magnetic properties, the FE structure results in a MF phase, whereas the PE structure remains only FM. For both cases, we consider two magnetization directions, ${\bm M} \parallel \hat{x}$ and ${\bm M} \parallel \hat{z}$ ($\parallel c$), within a Cartesian reference frame aligned with the Eu-O bonds in the PE structure. This yields four systems, called FM-I, FM-II, MF-I, and MF-II, each with a distinct magnetic space group (Fig.~\ref{fig:para_vs_ferro}). Unlike the RS phase with negligible MCA, the PE and FE structures exhibit a MCA energy of $E_{\mathrm{MCA}} = E({\bm M}\parallel a/b) - E({\bm M}\parallel c) \approx -0.1 \, \mathrm{meV}/\mathrm{f.u.}$, indicating that ${\bm M}$ has a slight preference for a direction perpendicular to the $c$-axis.

\begin{figure*}[ht!]
\centering
\includegraphics[width=1.0\linewidth]{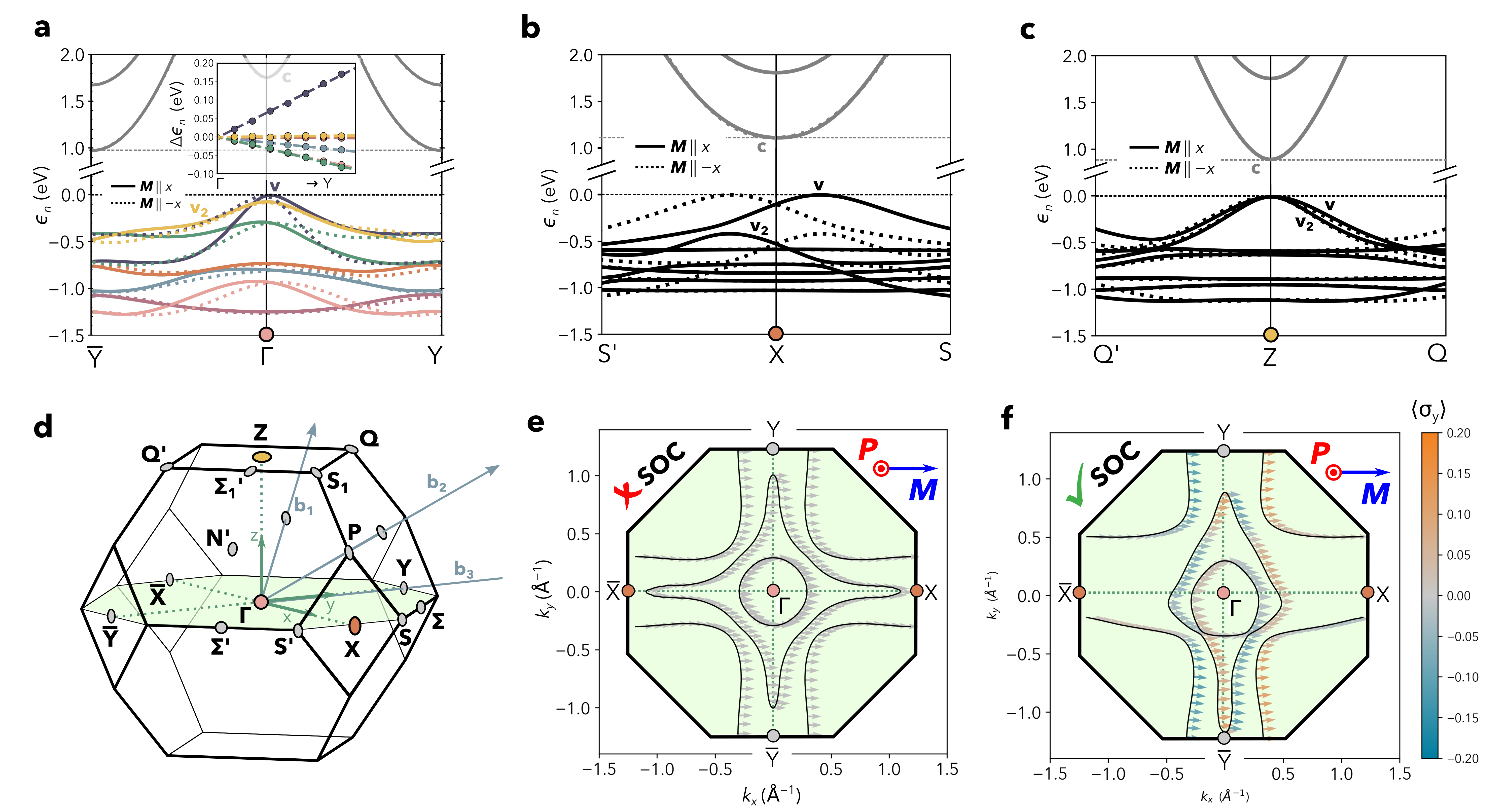}
\caption{
Band structure of MF-I. (a) Top valence bands along $\overline{\mathrm{Y}}-\Gamma-\mathrm{Y}$. Continuous (dotted) lines correspond to ${\bm M}\parallel (-)\hat{x}$. Inset: linear-in-$k$ behavior of $\Delta \epsilon_n$. (b) Band structure around the X. (d) Brillouin zone. (e) Constant energy cut at $E-E_\mathrm{F}=-0.4$ eV in the $k_z=0$ plane without SOC. (f) Same as (e) with SOC, also showing the spin polarizations. }
\label{fig:bands} 
\end{figure*}

The band structure of FM-I and FM-II display full inversion symmetry with respect to all special points in the BZ (see Fig.~\ref{fig:para_vs_ferro}-a and -b for a schematic and Fig.~S1-a and -b in the SI for the DFT results). In contrast, the band structure of MF-I (${\bm M} \perp \bm{P}$) is strikingly different.
For instance, bands show off-centering and asymmetric dispersion, $\epsilon_n({\bm k}) \neq\epsilon_n(-{\bm k})$, along the $\overline{\mathrm{Y}}-\Gamma-\mathrm{Y}$ direction, for ${\bm k}=(0,k_y,0)$, as shown in Fig.~\ref{fig:bands}-a. This effect is particularly pronounced for the Eu $4f$ valence bands, where the asymmetry $\Delta \epsilon_n ({\bm k}) = \epsilon_n({\bm k}) - \epsilon_n(-{\bm k})$ varies linearly for small $k_y$ (inset of Fig.~\ref{fig:bands}-a), with a maximum of $\approx 330$ meV. 

A similar off-centering and asymmetric dispersion of the top valence bands with respect to the $k_y\leftrightarrow-k_y$ inversion is also observed at other special points in the BZ of MF-I (Fig.~\ref{fig:bands}-d). The effect is therefore widespread, extending throughout the entire BZ and not limited to its center. For example, it appears around $\mathrm{Z}=(0.5, 0.5, -0.5)$ (in reciprocal lattice units) (Fig.~\ref{fig:bands}-c) and around $\mathrm{X}=(-0.5, 0.5, 0)$ (Fig.~\ref{fig:bands}-b), where it is even more pronounced than at $\Gamma$. Importantly, this band asymmetry is not limited to the valence bands, although it is most pronounced for them. 

In MF-II (${\bm M} \parallel \bm{P}$), the bands are symmetric along $\overline{\mathrm{Y}}-\Gamma-\mathrm{Y}$. Band asymmetry is observed around $\mathrm{X}$, but only along the low-symmetry lines in the BZ, with the maximal $\Delta \epsilon_n$ an order of magnitude smaller than in MF-I (see Fig. S2 in the SI).

{\it Symmetry analysis. }The NRBE is an intrinsic property of the MF phase as it requires both non-zero $\bm P$ and $\bm M$, but is enabled by SOC, which couples spin, momentum, and lattice, making MF-I distinct from MF-II. The role of SOC can be addressed using magnetic space groups, whose symmetry elements other than inversion and time reversal determine the different NRBE realizations (see also the End Matter). MF-I and MF-II belong to type-III magnetic space groups $Im'm2'$ (${\bm M} \perp \bm{P}$) and $I4m'm'$ (${\bm M} \parallel \bm{P}$), respectively. 
At lowest order in $\bm k$ within $\bm k\cdot\bm p$ perturbation theory, the effective Hamiltonians at $\Gamma$ read:
\begin{eqnarray}
H_{\mathrm{MF-I}}&=&\epsilon_n+\alpha_{0n}k_y + (B_x-\alpha_{1n} k_y)\sigma_x+\alpha_{2n} k_x\sigma_y,\nonumber\\
H_{\mathrm{MF-II}} &=&\epsilon_n+B_z\sigma_z\,+\,\alpha_{Rn} \left(k_x\sigma_y - k_y\sigma_x\right).\label{eq:model}
\end{eqnarray}
Here $\epsilon_n$ is the spin-independent energy of the $n$-th band, $ \sigma_{x,y}$ are the Pauli matrices, $\bm B$ is the effective Zeeman magnetic field accounting for the exchange splitting, and $\alpha_{in}$ parameterize the SOC-induced momentum-dependent terms. 
The band dispersion of MF-I is linear in $k_y$ (i.e., along $\bm P\times\bm M$), with the NRBE quantified as $\Delta \epsilon_{n}(0,k_y)=2 [\alpha_{0n}+\mbox{sgn}(B_x)\alpha_{1n}] k_y$ for $B_x\gg\alpha_{in}$, consistent with the DFT results (inset of Fig.~\ref{fig:bands}-a). Notably, the NRBE reverses when ${\bm M}$ is flipped to $-{\bm M}$ or ${\bm P}$ to ${-\bm P}$, as depicted by dotted lines in Fig.~\ref{fig:bands}-a, since all $\alpha_{in}$ are odd under inversion, while $\bm B$ and $\alpha_{0n}$ are odd under time reversal. Additionally, SOC induces a spin-$y$ polarization orthogonal to $\bm{M}$, vanishing only at $k_x = 0$, in agreement with DFT. Fig.~\ref{fig:bands}-e and -f show the 2D spin-texture in the $k_z=0$ plane calculated by DFT at the constant energy surface $E-E_\mathrm{F}=-0.4$ eV cutting through the spin-majority valence bands. Without SOC, the NRBE is absent, and there is only a spin-$x$ polarization (gray arrows in Fig.~\ref{fig:bands}-e). With SOC, the NRBE appears, together with a small spin-$y$ polarization, which vanishes along the $\overline{\mathrm{Y}}-\Gamma-\mathrm{Y}$ line (spin-$y$ polarization values are represented using the color scale in Fig.~\ref{fig:bands}-f). 
For MF-II, the band dispersion is instead quadratic, displaying no NRBE. Higher-order odd terms are forbidden at special points with $4m'm'$ little group, such as $\Gamma$ and $\mathrm{Z}$. Nonetheless, at $\mathrm{X}$, the lower $2m'm'$ symmetries allow a $k_xk_yk_z$ term, enabling NRBE, consistent with DFT results. 

\begin{figure}[ht!]
\centering
\includegraphics[width=1.0\linewidth]{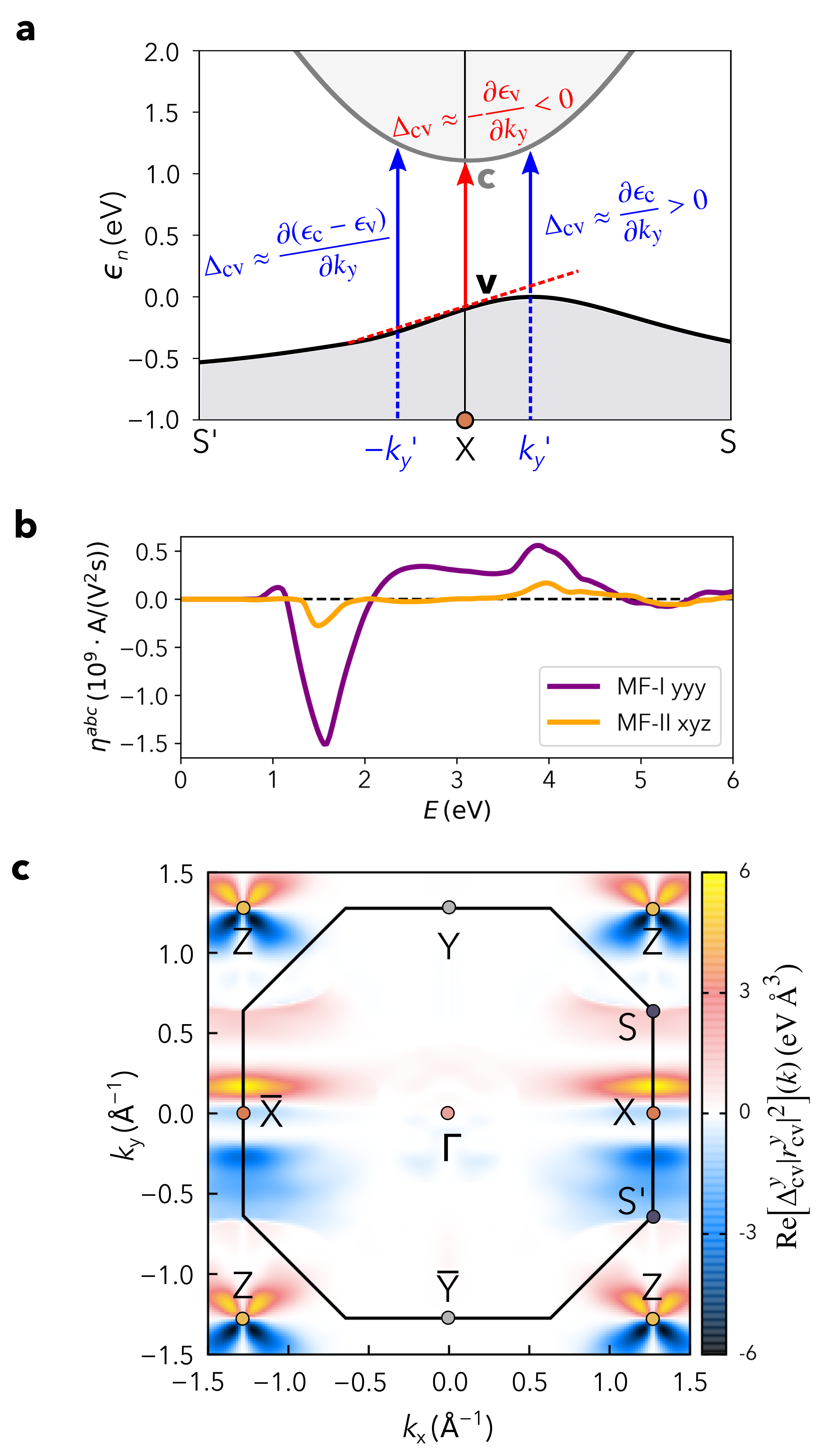}
\caption{Injection photoconductivity. (a) Schematic of the band velocity weight factor for MF-I. (b) $\eta_\mathrm{S}^{yyy}$ and $\eta_\mathrm{S}^{xyz}$ for MF-I and MF-II. (c) Integrand of $\eta^{yyy}$, corresponding to v-c electronic transitions in the $k_z=0$ plane of the BZ.}
\label{fig:conductivity} 
\end{figure}

{\it Photoconductivity.} Band asymmetry induces non-reciprocal charge transport \cite{Tokura2018}. In metals, this manifests as electrical magnetochiral anisotropy \cite{Rikken2005}, where electrical resistance depends on the charge current direction \cite{Ideue2017}. In the semiconducting multiferroic EuO, it is related to the BPE under linearly polarized light. 

The intrinsic BPE in multiferroics comes from two primary contributions: \textit{the shift current}, associated with the dc component of the second-order response to the light's dipole electric field \cite{Baltz1981,Sipe2000,Nastos2012,Young2012,Ibanez2018,Uriona2023}, and \textit{the injection current} \cite{Aversa1995,Sipe2000,Nastos2012,Uriona2023}. As shown in studies on bilayer magnetic systems \cite{Zhang2019,Wang2020,Zhang2022}, it is the injection current in particular that is sensitive to band inversion asymmetries, allowing for the detection of the NRBE.

The injection photocurrent density driven by an electric field $\mathbfcal{E}(\omega)$ of frequency $\omega$ is $j^a= \tau \sum_{bc}\eta^{abc} (\omega) [\mathcal{E}^b(\omega)\mathcal{E}^c(-\omega)]$
where $a,b$, and $c$ ($=x,y,z)$ label the vectors' and tensor components, $\tau$ is the carriers' relaxation time, and \cite{Sipe2000,Uriona2023}
\begin{equation}
\eta^{abc}(\omega) = -\frac{\pi e^3}{\hbar^2}\int\frac{d^3 \bm{k}}{(2\pi)^3}\sum_{nm}f_{nm} \Delta^{a}_{nm} r_{nm}^b r_{mn}^c\, \delta(\omega_{mn}-\omega)
\label{eq:eta}
\end{equation}
is a component of the third-rank injection photoconductivity tensor. The integrand in Eq. (\ref{eq:eta}) consists of the product of dipole matrix elements $r^{a}_{nm}$, weighted by the band velocity difference, $\Delta^{a}_{nm} = \frac{\partial (\epsilon_n - \epsilon_m)}{\partial k_a}$. The dipole effect becomes most prominent for transitions between the top valence (v) and bottom conduction (c) bands near the band gap minima (see section S3-E in the SI), owing to the relation $r_{\mathrm{vc}} = v_{\mathrm{vc}}/(\epsilon_\mathrm{v}-\epsilon_{\mathrm{c}})$ \cite{Ibanez2018}. The band velocity weights are related to the NRBE, as illustrated in Fig. \ref{fig:conductivity}-a for the $\mathrm{S}' - \mathrm{X} - {\mathrm{S}}$ direction in the BZ of MF-I. First, the off-centering of the v band edge gives $-\frac{\partial \epsilon_\mathrm{v} }{\partial k_y}<0$ at X (red arrow) and $\frac{\partial \epsilon_\mathrm{c} }{\partial k_y}>0$ along the X-S line (blue arrow). Second, the pronounced asymmetry in the band dispersion results in unequal weights for $\frac{\partial \epsilon_\mathrm{v}}{\partial k_y}$ and $-\frac{\partial \epsilon_\mathrm{v}}{\partial k_y}$ along $k_y$ and $-k_y$, respectively. This prevents their cancellation, leading to an imbalance in photocarriers with opposite momenta, and producing a band-edge photocurrent along the $y$ direction.

To validate this analysis, we compute $\eta^{abc}(\omega)$ following Ref. \cite{Uriona2023} and use the Wannier-interpolation scheme \cite{Marzari1997,Mostofi2008} of the DFT band structure via the {\sc Wannier90} package \cite{Pizzi2020} (see section S3-C in the SI).
Since $\eta^{abc}(\omega)$ has no definite parity under the $b\leftrightarrow c$ exchange and is odd under time-reversal symmetry \cite{Zhang2019}, we only consider the symmetric part $\eta_S^{abc}=(\eta^{abc}+\eta^{acb})/2$, describing the response to linearly polarized light (the antisymmetric part is discussed in section S3-F of the SI). 
Consistent with the magnetic symmetry requirements, the calculations yield five independent components for MF-I, namely $\eta_\mathrm{S}^{yxx}$, $\eta_\mathrm{S}^{yzz}$, $\eta_\mathrm{S}^{xyx}$, $\eta_\mathrm{S}^{zyz}$, and $\eta_\mathrm{S}^{yyy}$, with the last being the largest. 
As shown in Fig. \ref{fig:conductivity}-b (purple line), $\eta_\mathrm{S}^{yyy}$ rises sharply from the onset energy $E=\hbar \omega\sim 1.2$ eV, reaching a maximum absolute value of $\sim -1.5\times10^{9}$ A/(V$^2$s) at $E\sim 1.6$ eV. 
In comparison, MF-II has a single independent component, $\eta_\mathrm{S}^{xyz}=-\eta_\mathrm{S}^{yxz}$ (orange line in Fig. \ref{fig:conductivity}-b), with a similar shape to $\eta^{yyy}$ in MF-I but much smaller in magnitude, consistent with the much weaker NRBE.

The integrand of $\eta^{yyy}$ for MF-I in Eq. (\ref{eq:eta}), including the most prominent transitions between v and c in the $k_{z} = 0$ plane of the BZ, as shown Fig.~\ref{fig:conductivity}-c. Hot spots appear around Z and X, which correspond to the point of band gap minimum, owing to the dipole effect. The hot spots around Z exhibit a symmetric flower-shaped structure, while those near X show significant asymmetry along $k_y$, qualitatively consistent with the expected effect of the band velocity term, as illustrated in Fig.~\ref{fig:conductivity}-a. However, the dipole matrix elements contribution is also quantitatively important (see section IV-E in the SI). Additionally, we note that the onset energy, $\sim 1.2$ eV for $\eta^{yyy}$ coincides with the band gap at X, while its maximum occurs at $E\sim 1.6$ eV, as the c band is highly dispersive along X$-$S. 

Finally, the BPE analysis is completed by including the shift current (see section S3-D in the SI). However, no $yyy$ shift photoconductivity component is found in MF-I. This implies that the current measured along the $y$ direction from light polarized in the same direction will exclusively have an injection contribution, providing a direct probe for the NRBE. Using the $\eta_\mathrm{S}^{yyy}$ peak value, the photocurrent density is $j^y\sim100$ mA/cm$^2$ for $\tau= 0.7$ ps \cite{Oliver1972} and an electric field of 1000 V/m. This far exceeds previously reported photocurrents in oxides \cite{Young2012,Young2012_2,Dai2023} and even in layered systems \cite{Zhang2018,Zhang2022}. Thus, the NRBE leads to a colossal BPE. Furthermore, we note that the current changes sign when switching either $\bm{P}$ or $\bm{M}$ (just like the NRBE), allowing for electrical and magnetic control.

{\it Conclusion.} We have investigated EuO in its multiferroic phase and predicted a giant, SOC-driven NRBE that can be controlled by varying the directions of $\bm{P}$ and $\bm{M}$. The emergence of this NRBE can be detected as an enhanced linear injection photocurrent $\bm{j} \parallel \bm{P} \times \bm{M}$, resulting in a colossal BPE.  Although our study addresses EuO, the large NRBE and related BPE could be quite general phenomena in multiferroics, potentially leading to novel cross-functionalities. \\

\section*{Acknowledgments}
A.D., B.Y, G.C. and S.P. acknowledge partial financial support by the Next-Generation-EU program via the PRIN-2022 SORBET (Grant No. 2022ZY8HJY), the ICSC initiative (National Center for High-Performance Supercomputing, Big Data and Quantum Computing), the PE4 Partenariato Quantum Science and Technology (NQSTI).
S.S. acknowledges financial support from the Vin\v{c}a Institute, provided by the Ministry of Education, Science, and Technological Development of the Republic of Serbia through the contract No. 451-03-136/2025-03/200017. S.S., G.C., A.D., and P.B. acknowledge additional funding from the Ministry of Foreign Affairs of Italy and the Ministry of Science, Technological Development, and Innovation of Serbia for the bilateral project ``Van der Waals Heterostructures for Altermagnetic Spintronics'', under the executive programme for scientific and technological cooperation between the two countries. J.I.-A. and A.R.P.-U. were supported by the European Union’s Horizon 2020 research and innovation programme under the European Research Council (ERC) grant agreement No. 946629, StG PhotoNow. Computational resources and support were provided by CINECA under the ISCRA initiative, specifically through the projects ISCRA-B HP10BA00W3 and ISCRA-C HP10C6WZ1O.

\section*{End Matter: Symmetry and effective models.}
Within a one-particle approximation, the  electronic properties of a magnetic system are described by a set of one-particle wave functions satisfying a self-consistent Schr\"odinger equation $H\psi_{n\bm k\sigma}=\epsilon_{n\sigma}(\bm k)\psi_{n\bm k\sigma} $, where the Hamiltonian can be decomposed as
\begin{eqnarray}
\label{eq:hamiltonian}
H &=& H_\mathrm{K}+H_\mathrm{Har}+ H_\mathrm{SOC}+H_\mathrm{ex}.
\end{eqnarray}
$H_\mathrm{K}$ and $H_\mathrm{Har}$ are the (spin-independent) kinetic and self-consistent Hartree potential terms, respectively, and $H_\mathrm{SOC}$ represents the SOC term. $H_\mathrm{ex}$ is the self-consistent Fock exchange term, which, in ferromagnets, causes a Zeeman-like splitting of the electronic bands for spin-up and spin-down electrons. $H_\mathrm{K}$ exhibits complete translational and rotational symmetry, while $H_\mathrm{Har}$ is invariant under all space group operations of the crystal. $H_\mathrm{SOC}$ is invariant under double space group operations and time-reversal symmetry. In contrast, the self-consistent $H_\mathrm{ex}$ distinguishes between up and down spins, thus explicitly breaking time-reversal symmetry, even though it does not fix the direction of the spin quantization axis with respect to crystalline directions. In order to remove this ambiguity when SOC is taken into account, a common approximation for ferromagnets consists in expressing the exchange term as
\begin{eqnarray}
    H_{ex} &=& \bm B\cdot\bm\sigma = -\frac{1}{2}\lim_{\mathcal{B}\to 0} \frac{\Delta E}{\vert\mathcal{B}\vert}\mathcal{B}\cdot\bm\sigma
\end{eqnarray}
where $\mathcal{B}$ is an auxiliary magnetic field having no other effect than fixing the spin quantization axis and $\Delta E$ is the Zeeman-like splitting between energy levels for spin-up and spin-down electrons \cite{Falicov1968}. Within this approximation, the Hamiltonian must be invariant under all magnetic space group operations, explicitly taking into account how time-reversal is broken and/or combines with translation/rotation symmetry elements. 

Using $\bm k\cdot\bm p$ perturbation theory, the band structure around a high-symmetry point $\bm k_0$ can be described by an effective spin-dependent Hamiltonian constrained by the symmetries of the little group of $\bm k_0$, that in magnetic systems comprise both unitary ($\bm u$) and antiunitary ($\bm a$) operations. Magnetic space groups $Im'm2'$ and $I4m'm'$ describing the symmetries of MF-I and MF-II are symmorphic, implying that any point in the BZ has a little group that is a subgroup of the magnetic point group. The strongest constraints on the form of the effective Hamiltonian are therefore found at $\Gamma$, displaying the full magnetic point group; any symmetry-allowed term existing at $\Gamma$ will also be allowed at other high-symmetry points, such as X and Z. 
The spin-momentum part of the effective $\bm k\cdot\bm p$ Hamiltonian can generally be written as $\bm \gamma (\bm k)\cdot\bm \sigma$, where $\bm\gamma(\bm k)$ is a polynomial function of the crystal momentum and $\bm \sigma$ are the Pauli matrices accounting for the spin degrees of freedom. 
Magnetic symmetry constraints can be expressed as $\bm u H(\bm k) \bm u^{-1}= H(\bm k)$ and $\bm a H(\bm k) \bm a^{-1}= H(-\bm k)$, or equivalently, by imposing that the $\bm \gamma (\bm k)\cdot\bm \sigma$ term is invariant under all symmetry operations using the transformation rules for crystal momentum, $\bm k$, and spin operators, $\bm \sigma$, provided in Tables \ref{table:chartab_MFI} and \ref{table:chartab_MFII} for MF-I and MF-II, respectively.

\begin{table}[h]
\begin{tabular}{c|p{3cm}p{3cm}c}
$C_{2v}(C_s)$ & \centering $\{k_x, k_y, k_z\}$ & \centering $\{\sigma_x, \sigma_y, \sigma_z\}$&\\
\hline
$1$ & \centering $\{k_x, k_y, k_z\}$ & \centering $\{\sigma_x, \sigma_y, \sigma_z\}$&\\
$m_{100}$ & \centering$\{-k_x, k_y, k_z\}$ & \centering$\{\sigma_x, -\sigma_y, -\sigma_z\}$&\\
$m'_{010}$ & \centering$\{-k_x, k_y, -k_z\}$ & \centering$\{\sigma_x, -\sigma_y, \sigma_z\}$&\\
$2'_{001}$ & \centering$\{k_x, k_y, -k_z\}$ & \centering$\{\sigma_x, \sigma_y, -\sigma_z\}$&
\end{tabular}
\caption{Transformation rules for $\bm k$ and $\bm\sigma$ under the symmetry operation of the magnetic point group $C_{2v}(C_s)$, relevant for MF-I, with $Im'm2'$ symmetries, which comprises the unitary elements $1$, $m_{100}$ and the antiunitary elements $m_{010}'$, $2'_{001}$. The prime symbol denotes the antiunitary time-reversal operation $\theta$.}\label{table:chartab_MFI}
\end{table}

\begin{table}
\begin{tabular}{c|
>{\centering\arraybackslash}p{3cm}
>{\centering\arraybackslash}p{3cm}c}
$C_{4v}(C_4)$ & $\{k_x, k_y, k_z\}$ & $\{\sigma_x, \sigma_y, \sigma_z\}$&\\
\hline
$1$              & $\{k_x, k_y, k_z\}$ & $\{\sigma_x, \sigma_y, \sigma_z\}$&\\
$2_{001}$        & $\{-k_x, -k_y, k_z\}$ & $\{-\sigma_x, -\sigma_y, \sigma_z\}$&\\
$4^+_{001}$      & $\{k_y, -k_x, k_z\}$ & $\{\sigma_y, -\sigma_x, \sigma_z\}$&\\
$4^-_{001}$      & $\{-k_y, k_x, k_z\}$ & $\{-\sigma_y, \sigma_x, \sigma_z\}$&\\
$ m'_{100}$      & $\{k_x, -k_y, -k_z\}$ & $\{-\sigma_x, \sigma_y, \sigma_z\}$&\\
$m'_{010}$       & $\{-k_x, k_y, -k_z\}$ & $\{\sigma_x, -\sigma_y, \sigma_z\}$&\\
$ m'_{110}$      & $\{k_y, k_x, -k_z\}$ & $\{-\sigma_y, -\sigma_x, \sigma_z\}$&\\
$m'_{\bar{1}10}$ & $\{-k_y, -k_x, -k_z\}$ & $\{\sigma_y, \sigma_x, \sigma_z\}$&
\end{tabular}
\caption{Transformation rules for $\bm k$ and $\bm\sigma$ under the symmetry operation of the magnetic point group $C_{4v}(C_4)$, relevant for MF-II, with $I4m'm'$ symmetries, which comprises unitary elements $1$, $4_{001}^+$, $4_{001}^-$, $2_{001}$ and the antiunitary elements $m'_{100}$, $m'_{010}$, $m'_{110}$, $m'_{\bar{1}10}$. The prime symbol denotes the antiunitary time-reversal operation $\theta$.}\label{table:chartab_MFII}
\end{table}

For MF-I, with $\bm B=(B_x,0,0)$, both $k_y$ and $\sigma_x$, as well as their product, are invariant under all magnetic symmetry operations. Likewise, the product $k_x\sigma_y$ is also invariant under the magnetic symmetry operations of the $m'm2'$ point group. In contrast, for MF-II, with $\bm B=(0,0,B_z)$, only the linear combination $k_x\sigma_y-k_y\sigma_x$ remains invariant under magnetic group operations. These constraints lead to the effective models of Eq. (\ref{eq:model}), which describe the well-known Rashba effect in the presence of a magnetic field perpendicular (MF-I) and parallel (MF-II) to the polar axis \cite{Ideue2017,Yamauchi2019,Tao2020}. The resulting dispersion of $n$-th band, where $n$ is the quantum number labeling the eigensolutions of the spin-independent part of Eq. (\ref{eq:hamiltonian}) is given by
\begin{eqnarray}
    \epsilon_{n,\pm}^{MF-I} (\bm k)&=& \epsilon_n+\alpha_{0n} k_y \pm \sqrt{(B_x-\alpha_{1n} k_y)^2+\alpha_{2n}^2 k_x^2}, \nonumber\\
    \epsilon_{n,\pm}^{MF-II} (\bm k)&=& \epsilon_n\pm\sqrt{B_z^2+\alpha_{Rn}^2(k_x^2+k_y^2)}.
\end{eqnarray}
The exchange-driven Zeeman-like field $\bm B$ is much larger than the Rashba SOC, separating majority- and minority-spin bands by $\sim 2 B_\alpha$. In this regime, the Rashba term does not introduce additional band splitting. It provides a correction to the band dispersion which is odd in $\bm k$ for MF-I, thereby leading to NRBE, while it is even in $\bm k$ for MF-II.  Nevertheless, it can still contribute to $\bm k$-dependent spin-polarization of the bands. For MF-I, the spin expectation values are given by

\begin{eqnarray}
    \langle \sigma_x\rangle_\pm &=& \pm\frac{B_x-\alpha_{1n} k_y}{\sqrt{(B_x-\alpha_{1n} k_y)^2+\alpha_{2n}^2k_x^2}},\nonumber\\
    \langle \sigma_y\rangle_\pm &=&\pm\frac{\alpha_{2n} k_x}{\sqrt{(B_x-\alpha_{1n} k_y)^2+\alpha_{2n}^2k_x^2}}.
\end{eqnarray}

For MF-II, the Rashba term leads to a spin-polarization component perpendicular to the ferromagnetic axis, exhibiting the characteristic Rashba spin-texture (i.e., tangential to the quasimomenta), and with an amplitude which scales roughly as $\sim \alpha_{Rn}/B_z$.

Using Table \ref{table:chartab_MFII}, it is straightforward to show that no other higher-order odd terms in $\bm k$ (whether spin-independent or coupled to the $\sigma_z$ component) are allowed by the magnetic point group $4m'm'$. This therefore rules out the possibility of realizing the NRBE around $\Gamma$ or Z for MF-II. In contrast, the little group of the X point is reduced to $C_{2v}(C_2)$ ($2m'm'$), consisting of the symmetry operations $1$, $2_{001}$, $m'_{100}$ and $m'_{010}$. Under this lower symmetry, the transformation rules still only permit the linear combination $k_x\sigma_y-k_y\sigma_x$. However, both $k_x k_y k_z$ and $(k_x k_y k_z)\sigma_z$ are invariant under the $2m'm'$ symmetry operations. This enable odd-in-$k$ band dispersions around X, thereby giving rise to the NRBE.

\renewcommand{\thesection}{S\arabic{section}}
\renewcommand{\theequation}{S\arabic{equation}}
\renewcommand{\thefigure}{S\arabic{figure}} 
\renewcommand{\thetable}{S\arabic{table}}

\onecolumngrid
\section*{Supporting Information} 
\addcontentsline{toc}{section}{Supplementary Information}

\setcounter{figure}{0} 
\setcounter{table}{0}  
\setcounter{section}{0} 
\setcounter{equation}{0} 

\section{Computational details for the DFT calculations}
\label{sec:comp_details}
DFT calculations are performed by using the Vienna {\it Ab-initio} Simulation Package ({\sc VASP})~\cite{Kresse1993_si,Kresse1996_si,Kresse1996b_si}. The Perdew-Burke-Ernzerhof (PBE) \cite{Perdew1996_si} generalized gradient approximation (GGA) is adopted for the exchange-correlation density functional in combination with a Hubbard ``$+U$'' correction. 
The structural optimization under a $-5\%$ biaxial compressive strain are carried out in the FM state within a collinear spin-polarized approach. A plane wave cutoff of 520 eV is used, and the Brillouin zone (BZ) is sampled with a $10 \times 10 \times 10$ $\bm{k}$-points grid. In contrast, for the four studied configurations presented in Fig.~1 of the Letter, the calculations are performed within the non-collinear framework, including spin-orbit coupling (SOC), and with a denser $16 \times 16 \times 16$ $\bm{k}$-points grid. 

The application of a Hubbard correction to the electronic structure of EuO is crucial, as previous DFT studies consistently show that both L(S)DA and GGA inaccurately predict a metallic ground state, in sharp contrast to experimental findings, which establish that paramagnetic EuO in the RS phase is a semiconductor with a band gap of 1.12 eV at room temperature \cite{An2011}.
DFT+$U$ correctly predicts the material to be a semiconductor. However, the results reported in the literature exhibit considerable quantitative variations, depending on the specific implementation, whether based on the schemes by Liechtenstein {\it et al.} \cite{Liechtenstein1995_si} or Dudarev {\it et al.} \cite{Dudarev1998_si}, as well as the chosen values for the Hubbard $U$ and Hund $J$ parameters, and the orbitals to which they are applied~\cite{Ghosh2004_si,Larson2006_si,Shi2008_si,Ingle2008_si,An2011_si}. Given that there is no consensus on the most reliable DFT+$U$ scheme for EuO, we adopted a heuristic approach, guided by experimental data, to select a suitable DFT+$U$ method for the scope of our work.

In the FM state, which is relevant for our study, the lowest conduction bands are spin-split by $\Delta E_{\uparrow \downarrow} \approx 0.6 \, {\rm eV}$. As a rule of thumb, the band gap measured in the paramagnetic (PM) state is expected to be reduced by half of that value in the FM state. Specifically, the spin-up band gap should decrease to approximately $0.8$ eV, while the spin-down band gap should increase to around $1.4$ eV. Recent studies, both theoretical (using reliable GW calculations \cite{An2011_si}) and experimental (via angle-resolved photoemission spectroscopy \cite{ColonSantana2012_si}), have shown that the band gap in ferromagnetic EuO is $0.95$ eV and that it is indirect, with the valence band maximum (VBM) at $\Gamma$ and the conduction band minimum (CBM) at the X point in the BZ.

For our calculations, we have selected the simplified DFT$+U$ approach by Dudarev {\it et al.} \cite{Dudarev1998_si} with a single Hubbard parameter of $U_\mathrm{eff}=U-J = 7.4$ eV applied to the Eu $4f$ orbitals, and verified that the calculated lattice constant and band structure match experimental data. Similar $U_\mathrm{eff}$ values were used in Refs.~\cite{Larson2006_si, Shi2008_si}. This choice of $U_\mathrm{eff}$ results in a lattice constant of $a_0 = 5.178 \, {\rm \AA}$ in the FM state, which is just $0.7\%$ larger than the experimental value of $a_{\rm exp} = 5.144 \, {\rm \AA}$ \cite{Eick1956_si}. Furthermore, our band structure calculations (FM state, without SOC) yielded an indirect band gap of 1.03 eV, located between the VBM at $\Gamma$ and CBM at X, in agreement with experiments. When SOC is included the band gap reduces to 0.90 eV, which is even closer to the experimentally determined gap of 0.95 eV. As both the crystal structure and the main features of electronic structure are well reproduced with $U_\mathrm{eff} = 7.4$ eV on the Eu $4f$ orbitals, we maintain this value for all our calculations. 

\section{DFT band structures}
\begin{figure}[h!]
\centering
\includegraphics[width=0.6\linewidth]{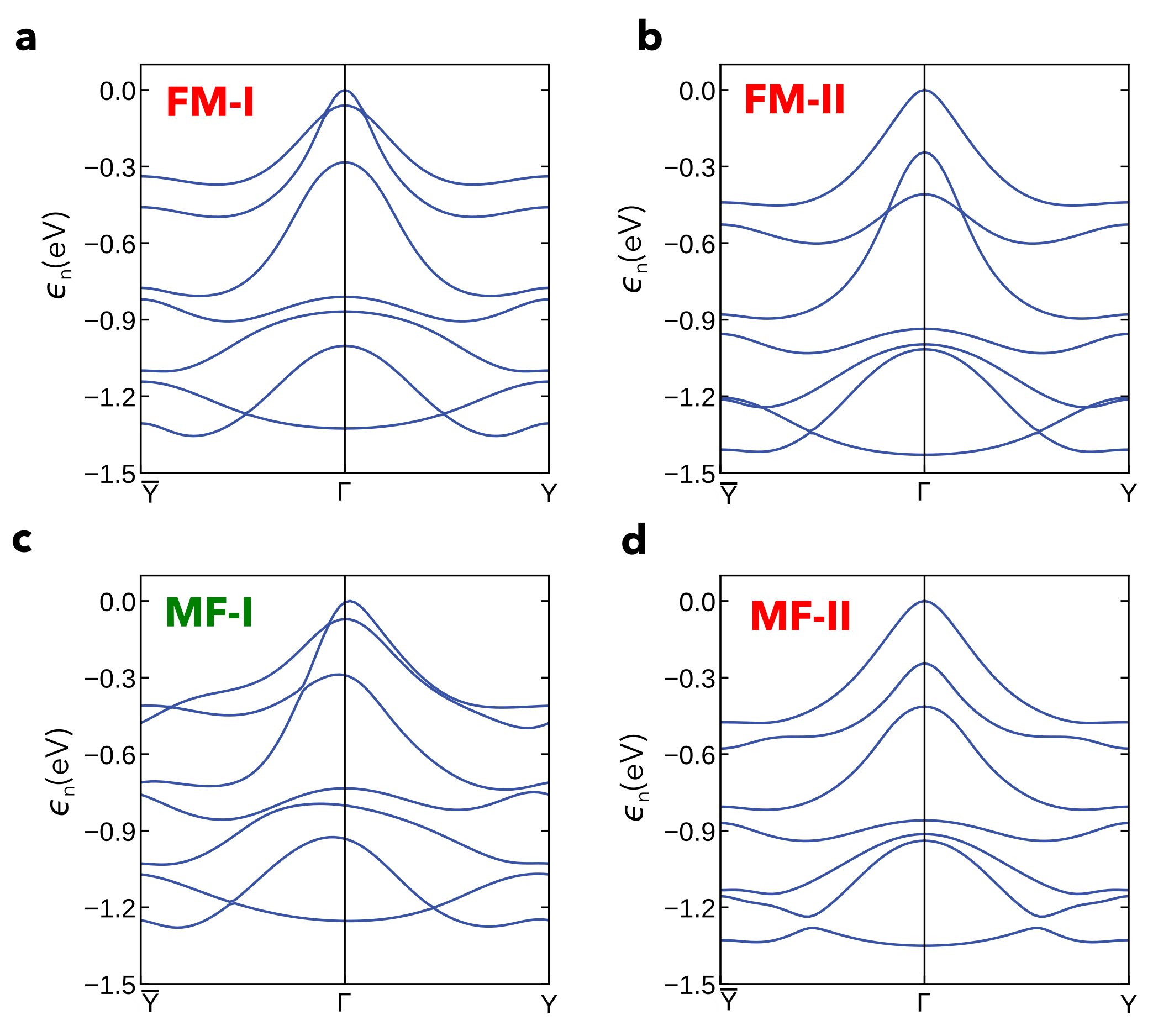}
\caption{Valence bands of the (a) FM-I, (b) FM-II, (c) MF-I, and (d) MF-II structures, which are depicted in Fig.~1 of the Letter.}
\label{fig:bands4} 
\end{figure}

In Fig.~\ref{fig:bands4}, we show the Eu $4f$ valence bands along the $\overline{\mathrm{Y}}-\Gamma-\mathrm{Y}$ path for the four structures in Fig. 1 of the Letter: FM-I, FM-II, MF-I, and MF-II, with the magnetization $\bm{M}$ either parallel or perpendicular to the $c$-axis (remember that $\bm{P} \parallel c$ in the MF phase). Although the band structures differ significantly, reflecting both structural differences between the FM and MF phases and variations in the direction of $\bm M$ within each phase, the $k_y \leftrightarrow -k_y$ asymmetry is observed exclusively in the MF-I configuration.

\begin{figure}[h!]
\centering
\includegraphics[width=1\linewidth]{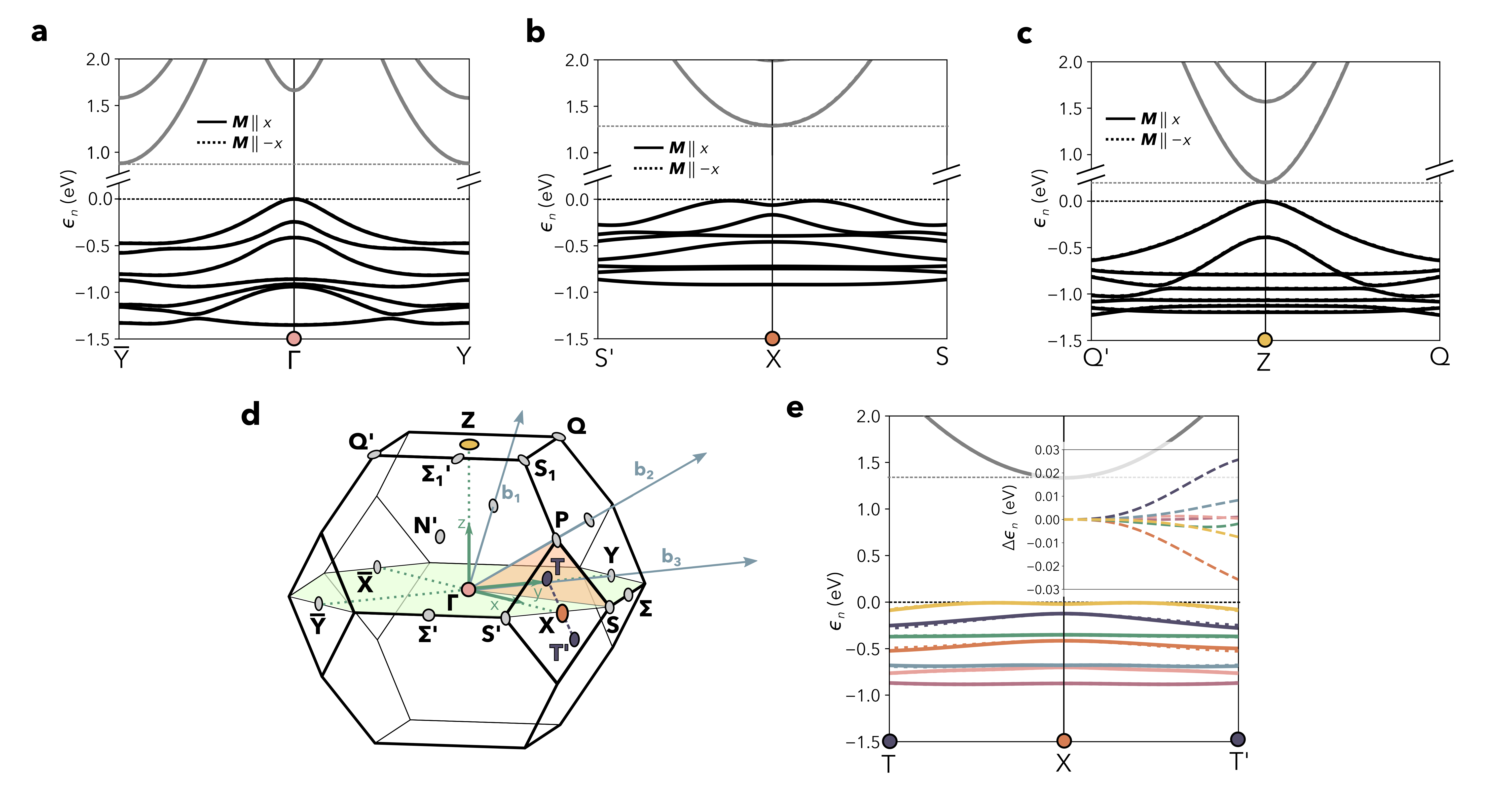}
\caption{The band structure of MF-II is symmetric along lines parallel to $k_y$ and around the special points $\Gamma$ (a), X (b), and Z (c) in the BZ. In contrast, an asymmetry emerges along low-symmetry lines, such as T$-$X$-$T$'$ (e), where T represents the center of the $\Gamma$PS triangle in the BZ and T$'$ is its mirror counterpart with respect to X (d).}
\label{fig:MFII-bands} 
\end{figure}

In Fig. \ref{fig:MFII-bands}, we present the band structure of MF-II (with ${\bm M} \parallel \bm{P}$) around various special points in the BZ, allowing direct comparison with the corresponding plots for MF-I (with ${\bm M} \perp \bm{P}$) shown in Fig. 2 of the Letter. Notably, unlike MF-I,
the MF-II bands are perfectly symmetric along  $\overline{\mathrm{Y}}-\Gamma-\mathrm{Y}$ (Fig.~\ref{fig:MFII-bands}-a), $\mathrm{S}' - \mathrm{X} - {\mathrm{S}}$ (Fig.~\ref{fig:MFII-bands}-b) and $\mathrm{Q}' - \mathrm{Z} - {\mathrm{Q}}$ (Fig.~\ref{fig:MFII-bands}-c). These paths are all parallel to $k_y$, supporting the conclusion from the Letter that the MF-II band structure shows no NRBE under $k_y \leftrightarrow -k_y$ reversal. However, MF-II can exhibit NRBE along other directions, with low symmetry, in the BZ. This becomes evident if we consider, for example, the path between the point T,  which is at the center of the $\Gamma$PS triangle in the BZ, and its mirror point T$'$ with respect to X (Fig.~\ref{fig:MFII-bands}-d). Along the path T$-$X$-$T$'$, the bands are asymmetric around X (Fig.~\ref{fig:MFII-bands}-e), but $\Delta \epsilon_n$ does not exceed 30 meV (inset of Fig.~\ref{fig:MFII-bands}-e), that is at least an order of magnitude smaller than $\Delta \epsilon_n$ around X in MF-I. Furthermore, $\Delta \epsilon_n$ around X is not linear in $k$. In fact, our analysis shows that the ${\bm k} \cdot {\bm p}$ energy bands exhibit an NRBE proportional to $k_x k_y k_z$ at the lowest order (see the symmetry analysis in the End Matter). 

\section{dc photocurrents}
 The bulk photovoltaic effect (BPE) comes from two main contributions: the shift and the injection currents.  In the following sections, we define these contributions and discuss results which complement those presented in the Letter.

\subsection{Shift current}\label{Sec.shift}

The shift photocurrent density driven by the light electric field $\mathbfcal{E}(\omega)$ of frequency $\omega$ is expressed as \cite{Sipe2000_si}
\begin{equation}\label{eq:shift_photocurrent}
j^a_{\text{shift}} = 2\sum_{bc}\sigma^{abc}(\omega)\text{Re}\left[\mathcal{E}^b(\omega)\mathcal{E}^c(-\omega)  \right], 
\end{equation}
where $a$, $b$, and $c$ are the indexes labeling the Cartesian directions, and $\sigma^{abc}$ is a component of the rank-three shift photoconductivity tensor. This tensor is symmetric under the $b\Leftrightarrow c$ exchange and even under time-reversal symmetry, so it transforms like the piezoelectric tensor. Denoting Bloch states of energy $\epsilon_n,\, \epsilon_m$ as $\ket{n},\, \ket{m}$, it can be shown that \cite{Sipe2000_si,Ibanez2018_si}
\begin{equation}\label{eq:shift_tensor_R}
\sigma^{abc}(\omega) = -\frac{i\pi e^3}{4\hbar^2}\int_{\text{BZ}}\frac{d^3 \bm{k}}{(2\pi)^3}\sum_{nm}f_{nm} \big[ r_{mn}^b r_{nm;a}^c
+ (b\Leftrightarrow c) \big] \times \big[\delta(\omega_{mn}+\omega)+\delta(\omega_{mn}-\omega)\big],
\end{equation}
where $e$ is the electron charge, $f_{nm} = f_n - f_m$ is the difference between the Fermi occupation factors, and $\omega_{nm} = \left(\epsilon_n - \epsilon_m\right)/\hbar$ is the optical excitation frequency. Additionally, we have introduced  
\begin{align}
r_{nm}^a & = (1-\delta_{nm})A_{nm}^a, \\
r_{nm;b}^a & = \partial_b r_{nm}^a -i\left(A_{nn}^b - A_{mm}^b \right)r_{nm}^a,
\end{align}
where $A_{nm}^a$ is the Berry connection,
\begin{equation}
A_{nm}^a = i\braket{n|\partial_a|m}.
\end{equation}
Note that, in these equations, the $\bm{k}$ dependence was omitted for concise notation, and $\partial_a$ stands for $\partial/\partial k_a$.

\subsection{Injection current}
\label{sec:inj_curr}

The injection photocurrent density driven by the light electric field $\mathbfcal{E}(\omega)$ of frequency $\omega$ is expressed as \cite{Sipe2000_si,Uriona2023_si}
\begin{equation}\label{eq:inj_photocurrent}
j^a_{\text{inj}} = \tau\sum_{bc}\eta^{abc}(\omega)\mathcal{E}^b(\omega)\mathcal{E}^c(-\omega).
\end{equation}
Here, $\tau$ is the relaxation time, and $\eta^{abc}(\omega)$ is a component of the injection current photoconductivity tensor, which, using the same notation as in Section \ref{Sec.shift}, is given by \cite{Aversa1995_si,Uriona2023_si}
\begin{equation}\label{eq:inj_tensor_full}
\eta^{abc}(\omega) = -\frac{\pi e^3}{\hbar^2}\int_{\text{BZ}}\frac{d^3 \bm{k}}{(2\pi)^3}\sum_{nm}f_{nm} \frac{\partial \omega_{mn}}{\partial k^a} r_{nm}^b r_{mn}^c \times \delta(\omega_{mn}-\omega).
\end{equation}
The injection photoconductivity tensor has no definite parity under the $b\Leftrightarrow c$ exchange and is odd under time-reversal symmetry. Eq. (\ref{eq:inj_tensor_full}) is the same as Eq. (2) in the Letter.

$\eta^{abc}$ can be separated into symmetric (S) and antisymmetric (A) parts under the $b\Leftrightarrow c$ exchange \cite{Wang2020_si,Uriona2023_si}, 
\begin{subequations}
\label{eq:inj_sym_asym_components}
\begin{align}
\eta^{abc}_\mathrm{S}(\omega) & = \frac{1}{2}\left[\eta^{abc}(\omega) + \eta^{acb}(\omega)\right],\label{eq:inj_sym}\\
\eta^{abc}_\mathrm{A}(\omega) & = \frac{1}{2}\left[\eta^{abc}(\omega) - \eta^{acb}(\omega)\right].
\label{eq:inj_asym}
\end{align}
\end{subequations}
$\eta^{abc}_\mathrm{A}$ and $\eta^{abc}_\mathrm{S}$ are purely imaginary and real, respectively. The antisymmetric part describes the response that is in magnitude but opposite in sign for left‐ and right-circularly polarized light, thereby vanishing for linearly polarized light. Conversely, the symmetric part describes the response to linearly polarized light. The photocurrents associated to $\eta^{abc}_\mathrm{S}$ and $\eta^{abc}_\mathrm{A}$ are called ``linear'' and ``circular'' injection currents or, alternatively ``magnetic'' and ``normal'' injection currents \cite{Wang2020_si}. In the case of EuO, we are mostly interested in $\eta^{abc}_\mathrm{S}$, which we simply call ``injection current'' in the Letter. However, for completeness, we also discuss $\eta^{abc}_\mathrm{A}$ at the end of Section \ref{sec.result_Ainjection}.

\subsection{Computational details for the photoconductivity calculations}

The shift and injection photoconductivity tensors are calculated using the approach developed by Puente-Uriona {\it et al.} \cite{Uriona2023_si}, which is based on the Wannier-interpolation scheme (``Wannierization'') \cite{Marzari1997_si,Mostofi2008_si} of the DFT band structure as implemented in the {\sc Wannier90} package \cite{Pizzi2020_si}.
The Wannierization process includes the Eu $4f$ states in the spin-up sub-sector, as well as the $6s$ and $5d$ orbitals, resulting in a total of 19 Wannierized bands. A $100 \times 100 \times 100$ $\bm{k}$-point grid is used for the calculations, and convergence is confirmed by observing identical results with a denser $\bm{k}$-point grid.

\begin{figure}[h!]
\centering
\includegraphics[width=1\linewidth]{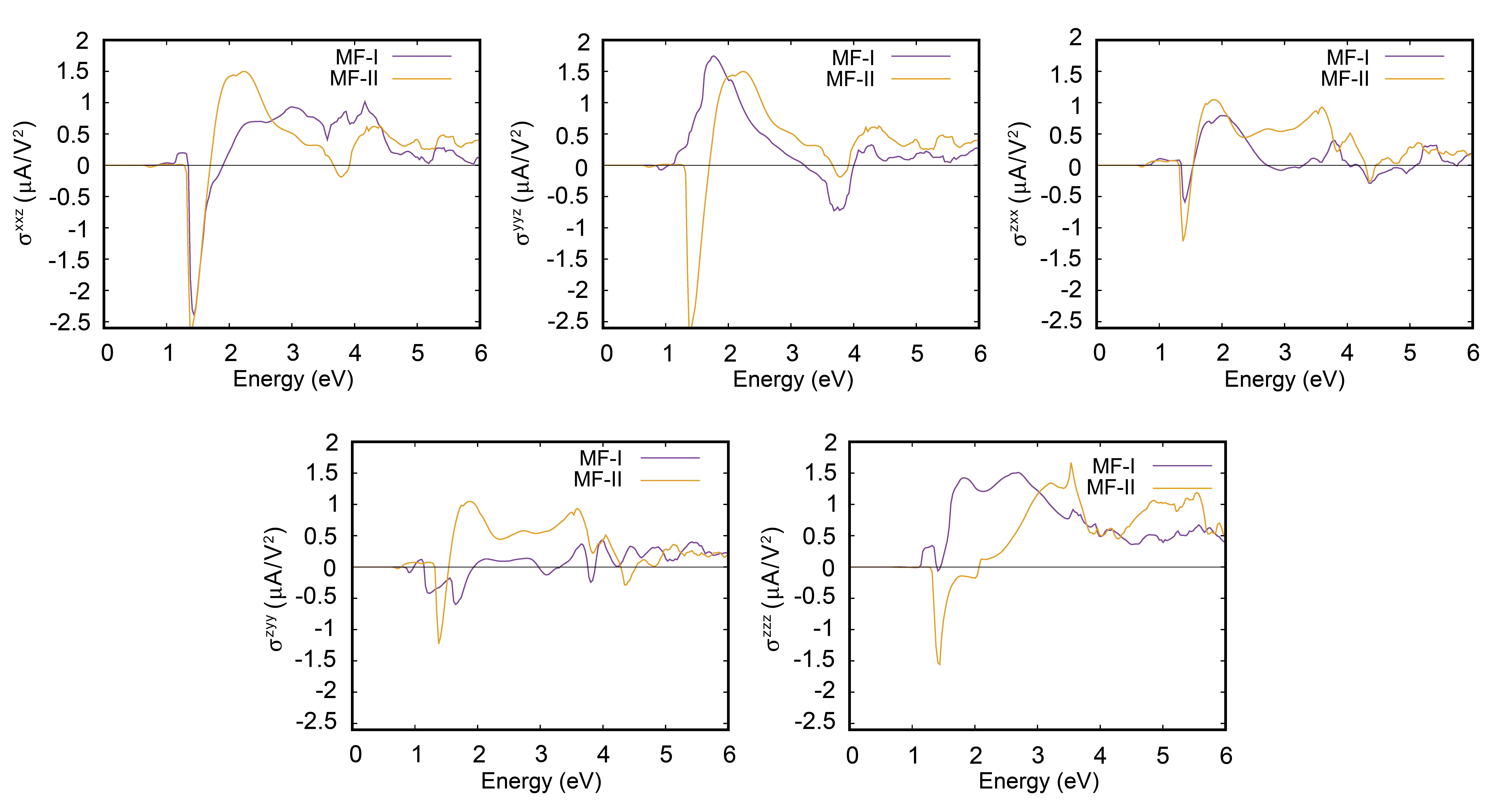}
\caption{Spectra of non-zero components of the shift photoconductivity tensor for MF-I ($\bm{M}\parallel \hat x$) and MF-II ($\bm{M}\parallel \hat z$).}
\label{fig:Shift} 
\end{figure}

\subsection{Results for the shift photoconductivity tensor}

The spectra of the non-zero components of the shift photoconductivity tensor for both the MF-I and MF-II configurations are presented in Fig. \ref{fig:Shift}. 

For MF-II $(\boldsymbol{M}\parallel \hat z)$, the $I4m'm'$ magnetic space group allows for five non-zero components of the shift photoconductivity tensor, with three of them being independent: $\sigma^{xxz}=\sigma^{yyz}$, $\sigma^{zxx}=\sigma^{zyy}$, and $\sigma^{zzz}$. In the visible energy range, $E=\hbar\omega\sim 1.5-3.3$ eV, $\sigma^{xxz}(=\sigma^{yyz})$ is the largest. It exhibits a sharp negative peak of approximately 2 $\mu$A$/$V$^2$ at $E\sim 1.5$ eV, changes sign to become positive at $E\sim 1.8$ eV, and shows a positive resonance centered around $2.1$ eV. A similar behavior is also found for the other components, although the sharp negative peak is slightly reduced compared to that seen in $\sigma^{xxz}$. Overall, the absolute magnitude of all non-zero components is comparable to that reported for the prototypical ferroelectric oxide material BaTiO$_3$ and PbTiO$_3$ in the same energy range \cite{Young2012_2_si}.

For MF-I $(\boldsymbol{M}\perp \hat z)$, the $Im'm2'$ magnetic space group allows for the same non-zero components of the shift current photoconductivity tensor as for MF-II. However, in MF-I, all components become independent: $\sigma^{xxz}$, $\sigma^{yyz}$, $\sigma^{zxx}$, $\sigma^{zyy}$, and $\sigma^{zzz}$. The spectrum of $\sigma^{xxz}$ exhibits minor changes in MF-I compared to MF-II, maintaining a sharp negative peak at the onset of the visible energy region ($E\sim 1.5$ eV), despite some redistribution of spectral weight at higher energies.  In contrast, the spectrum of $\sigma^{zxx}$, and more so the spectrum of $\sigma^{zyy}$, is significantly suppressed across the entire visible range. Finally, in the spectra of $\sigma^{yyz}$ and $\sigma^{zzz}$, the sharp negative peak observed for MF-II disappears in MF-I, while the positive resonance shifts toward lower energies. In the case of $\sigma^{yyz}$, this resonance also broadens.

The spectra of the components of the shift photoconductivity tensor flip in sign, $\sigma^{abc}(\omega)\rightarrow -\sigma^{abc}(\omega)$ when the electrical polarization, $\bm{P}$, is switched from $\hat z$ to $-\hat z$ for both MF-I and MF-II. In contrast, the spectra remain unchanged if the direction of the magnetization, $\bm{M}$, is reversed as the shift photoconductivity tensor is even under time reversal symmetry.

\begin{figure}[h!]
\centering
\includegraphics[width=1\linewidth]{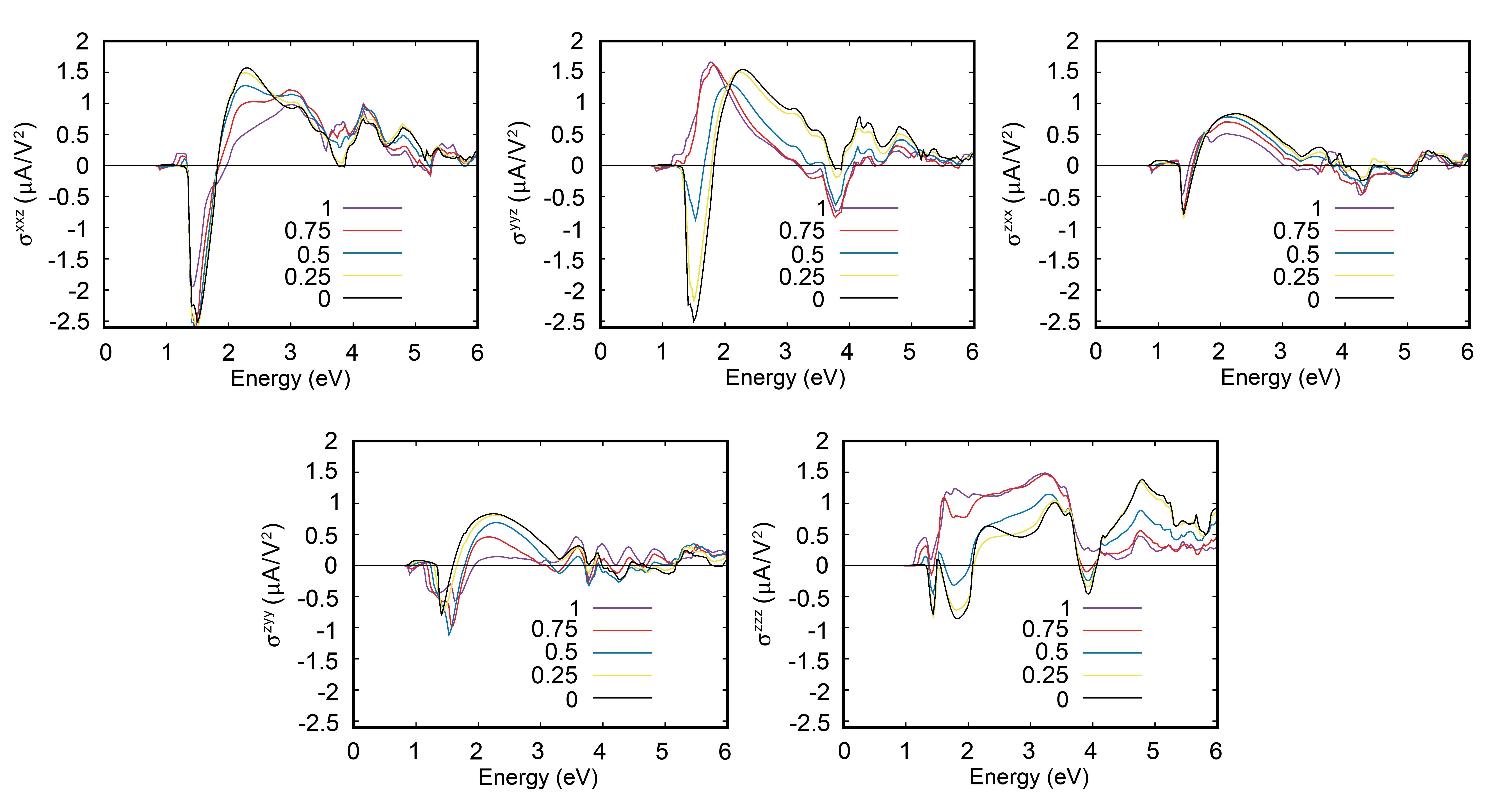}
\caption{Spectra of the non-zero components of the shift photoconductivity tensor of MF-I ($\bm{M}\parallel a$) as a function of re-scaled SOC strength. The re-scaling factor varies from $1$ (i.e., no re-scaling) to 0 (i.e., vanishing SOC).}
\label{fig:ShiftSOC} 
\end{figure}

Finally, we also analyze the dependence on SOC of the shift photoconductivity. Specifically, in Fig. \ref{fig:ShiftSOC}, we plot the spectra of the non-zero components for MF-I as a function of the re-scaled SOC strength. The rescaling factor varies from $1$ (i.e., no re-scaling) to 0 (i.e., vanishing SOC). We observe that there is a significant change of all components as the SOC strength is reduced,
eventually converging to the spectra observed in MF-II, where $\sigma^{xxz}=\sigma^{yyz}$, $\sigma^{zxx}=\sigma^{zyy}$, since MF-I and MF-II are identical without SOC.      

\begin{figure}[h!]
\centering
\includegraphics[width=1\linewidth]{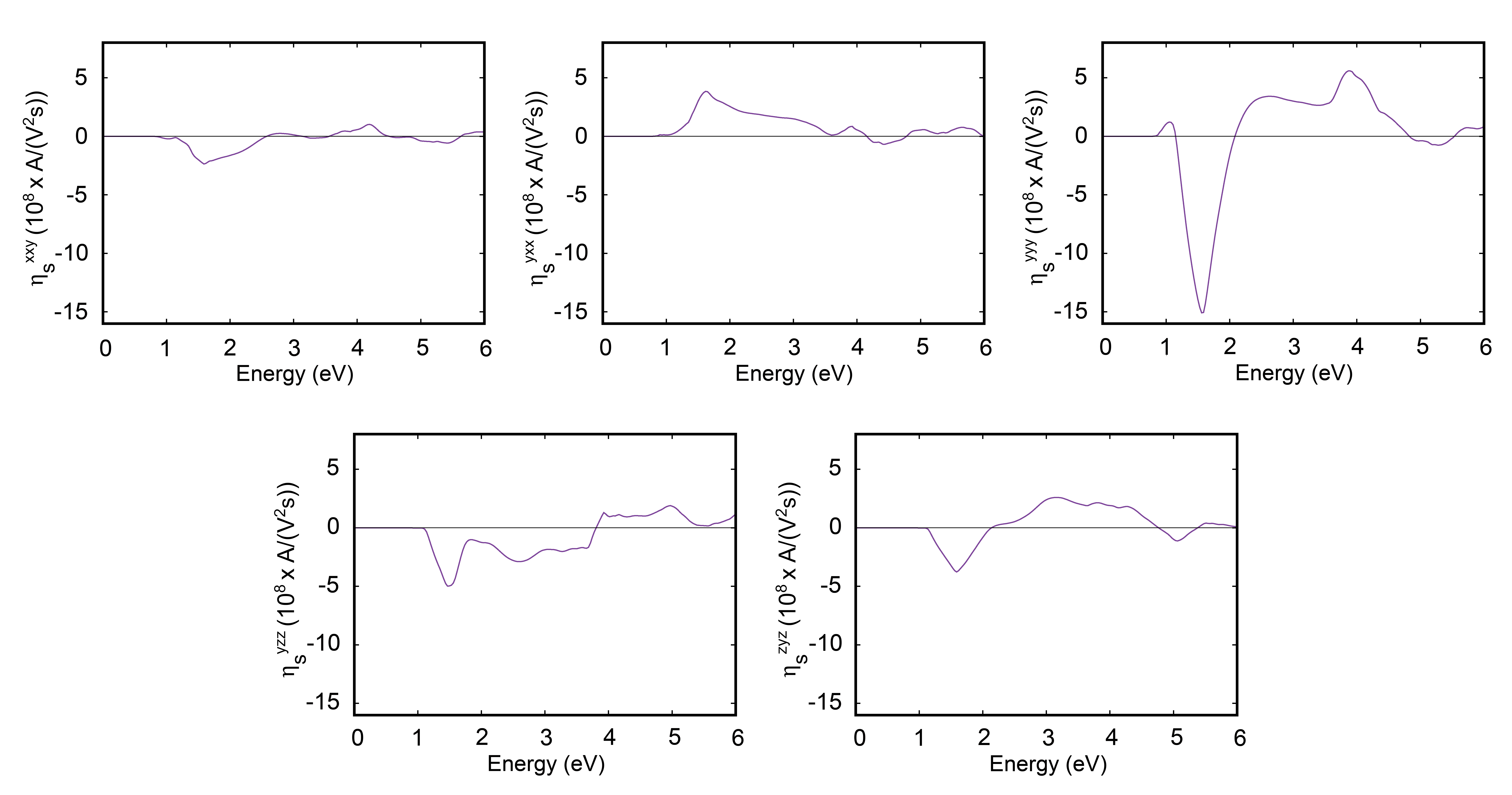}
\caption{Spectra of the non-zero components of the symmetric part of the injection current photoconductivity tensor of MF-I ($\bm{M}\parallel \hat x$).}
\label{fig:LinearInjection} 
\end{figure}

\subsection{Results for the symmetric part of the injection  photoconductivity tensor}\label{sec.result_Sinjection}

\subsubsection*{MF-I structure}
The spectra of the non-zero and independent components of the symmetric part of the injection photoconductivity tensor are presented in Fig. \ref{fig:LinearInjection} for MF-I $(\bm{M}\parallel a)$. 

As discussed in the Letter, there are seven non-zero components, five of which are independent: $\eta_\mathrm{S}^{yxx}$, $\eta_\mathrm{S}^{yyy}$, $\eta_\mathrm{S}^{yzz}$, $\eta_\mathrm{S}^{xyx}=\eta_\mathrm{S}^{xxy}$, and $\eta_\mathrm{S}^{zyz}=\eta_\mathrm{S}^{zzy}$. In the visible energy region, $\eta_\mathrm{S}^{yzz}$ is positive, while $\eta_\mathrm{S}^{xxy}$, $\eta_\mathrm{S}^{yzz}$, and $\eta_\mathrm{S}^{zyz}$ are negative, with maximum absolute values between $\sim3$ and $\sim5$ $\times10^{8}$ A$/($V$^2$s) at $E\sim 1.6$ eV. In contrast, the component $\eta_\mathrm{S}^{yyy}$, which corresponds to a current and an electric field perpendicular to both $\bm{P}$ ($\parallel \hat z$) and $\bm{M}$ ($\parallel \hat x$), displays a sharp and significantly higher negative peak, reaching an absolute value of $\sim15\times10^{8}$ A$/($V$^2$s) (also compare Fig. 4 in the Letter). This peak is an order of magnitude greater than those for the other components. As outlined in the Letter, such enhancement is a signature of the NRBE.   

\begin{figure}
\centering
\includegraphics[width=0.4\linewidth]{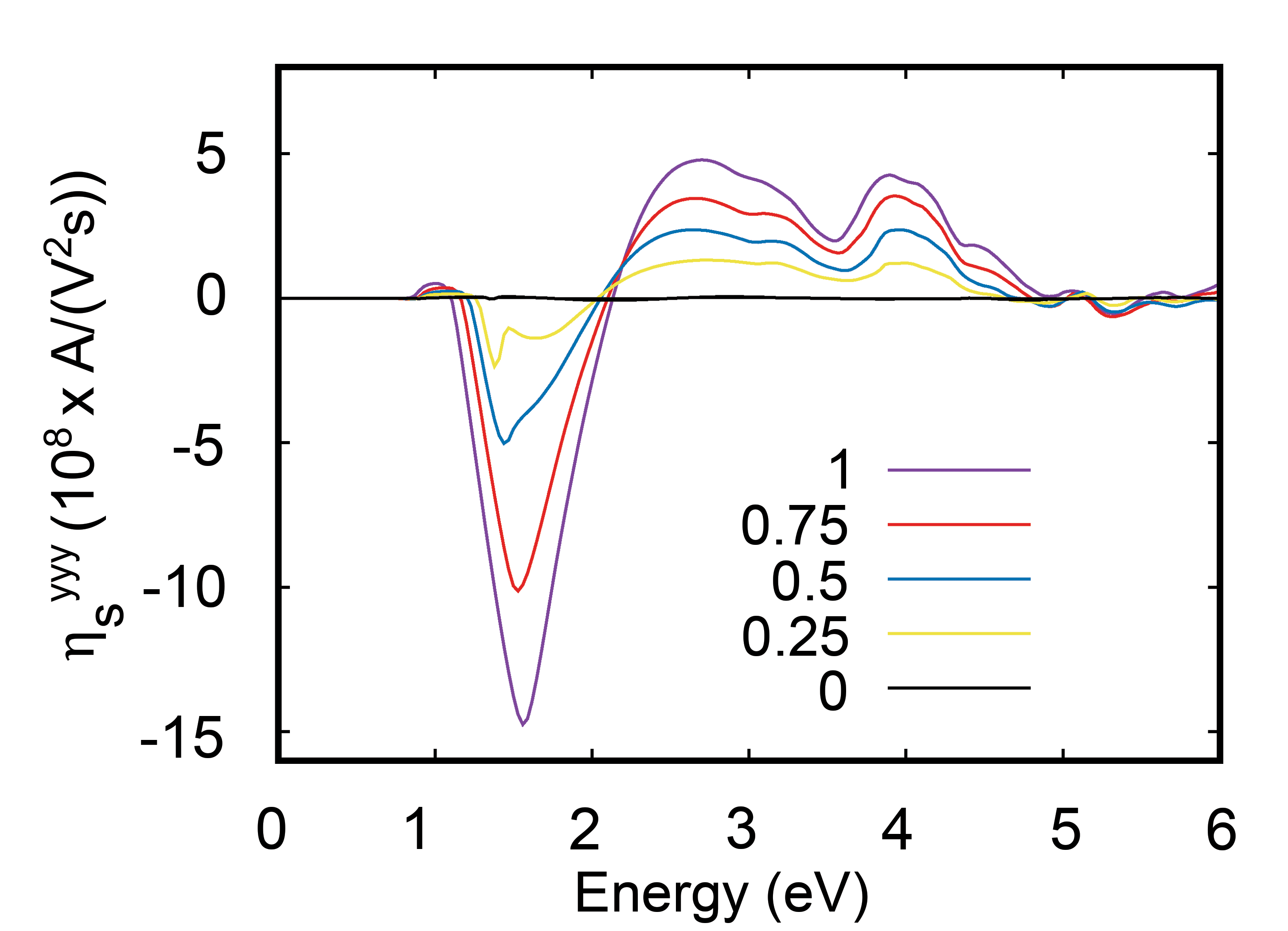}
\caption{Spectra of the component, $\eta^{yyy}_\mathrm{S}$, of the symmetric part of injection photoconductivity tensor of MF-I ($\bm{M}\parallel \hat x$) as a function of re-scaled SOC strength. The rescaling factor varies from $1$ (i.e., no re-scaling) to 0 (i.e, vanishing SOC).}
\label{fig:LinearInjectionSOC} 
\end{figure}

The spectra of the components of the linear injection photoconductivity tensor flip in sign, $\eta_\mathrm{S}^{abc}(\omega)\rightarrow -\eta_\mathrm{S}^{abc}(\omega)$, when either $\bm{P}$ is switched from $\hat z$ to $-\hat z$ or $\bm{M}$ is switched from $\hat x$ to $-\hat x$. In contrast, they remain unchanged if both $\bm{P}$ and $\bm{M}$ are reversed simultaneously. Furthermore, they vanish when SOC is turned off, as shown in Fig. \ref{fig:LinearInjectionSOC}  for the specific case of $\eta_\mathrm{S}^{yyy}$. This is because, without SOC, the $\bm{M}$ and $-\bm{M}$ configuration can not be distinguished and therefore $\eta_\mathrm{S}^{yyy}(\bm{M})=-\eta_\mathrm{S}^{yyy}(-\bm{M})=0$.

\begin{figure}[h!]
\centering
\includegraphics[width=0.6\linewidth]{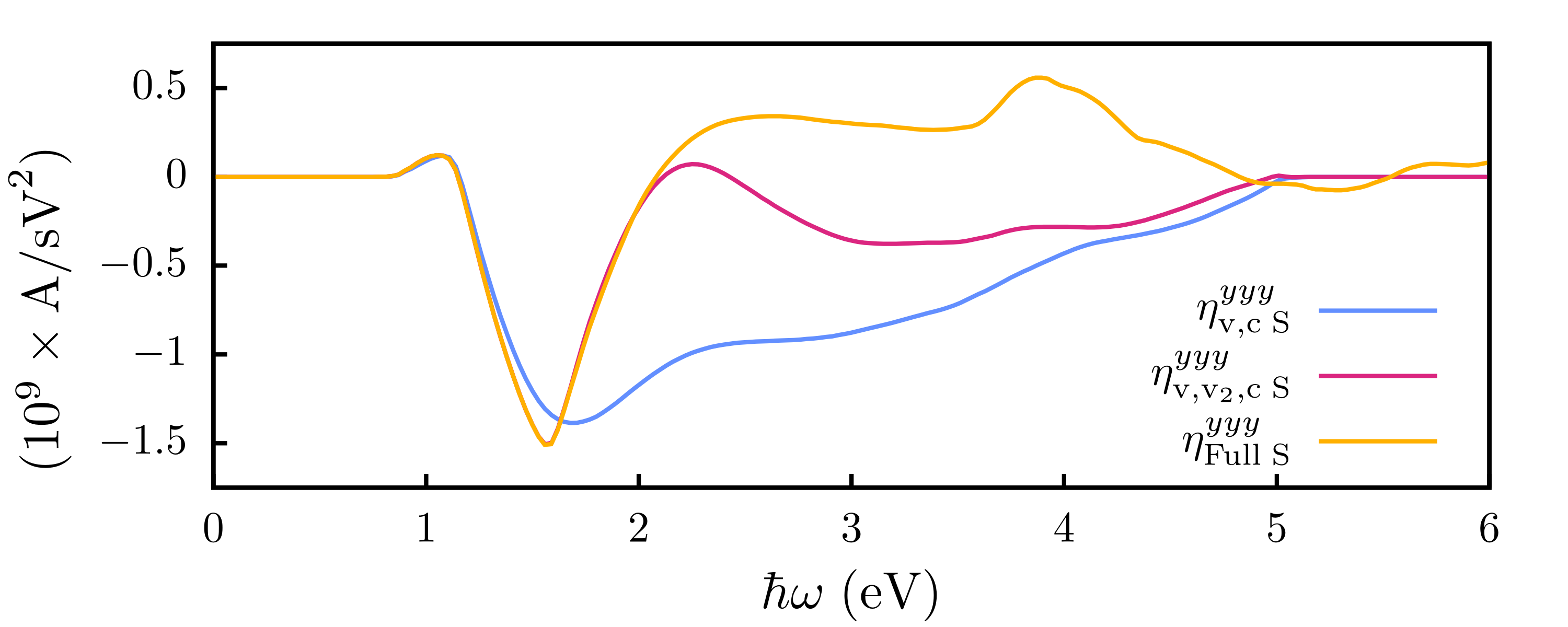}
\caption{Spectra of the symmetric component, $\eta^{yyy}_\mathrm{S}$, of the injection photoconductivity tensor of MF-I ($\bm{M}\parallel \hat{x}$) including only transitions between the $\text{v}$ and $\text{c}$ bands ($\eta^{yyy}_{\text{v},\text{c} \; \mathrm{S}}$, blue), the $\text{v}$, $\text{v}_2$ and $\text{c}$ bands ($\eta^{yyy}_{\text{v},\text{v}_{2},\text{c} \; \mathrm{S}}$, magenta) and all valence and conduction bands ($\eta^{yyy}_{\mathrm{Full} \; \mathrm{S}}$, yellow).}
\label{fig:injection_3sets} 
\end{figure}

To further analyze the origin of the injection current, in Fig.~\ref{fig:injection_3sets}, we show the contribution to $\eta^{yyy}_\mathrm{S}$ from transitions between the top valence band, $\text{v}$, and the bottom conduction band, $\text{c}$. Comparing this with the calculation that includes all bands demonstrates that the major contribution to the main peak around 1.6 eV originates exclusively from these transitions, justifying the analysis in the Letter based on these two states. Fig. \ref{fig:injection_3sets} also shows that including the second valence band $\text{v}_2$ significantly impacts the spectrum at energies above the main peak, but has a much smaller effect at energies around 1.6 eV.

\begin{figure}[h!]
\centering
\includegraphics[width=0.6\linewidth]{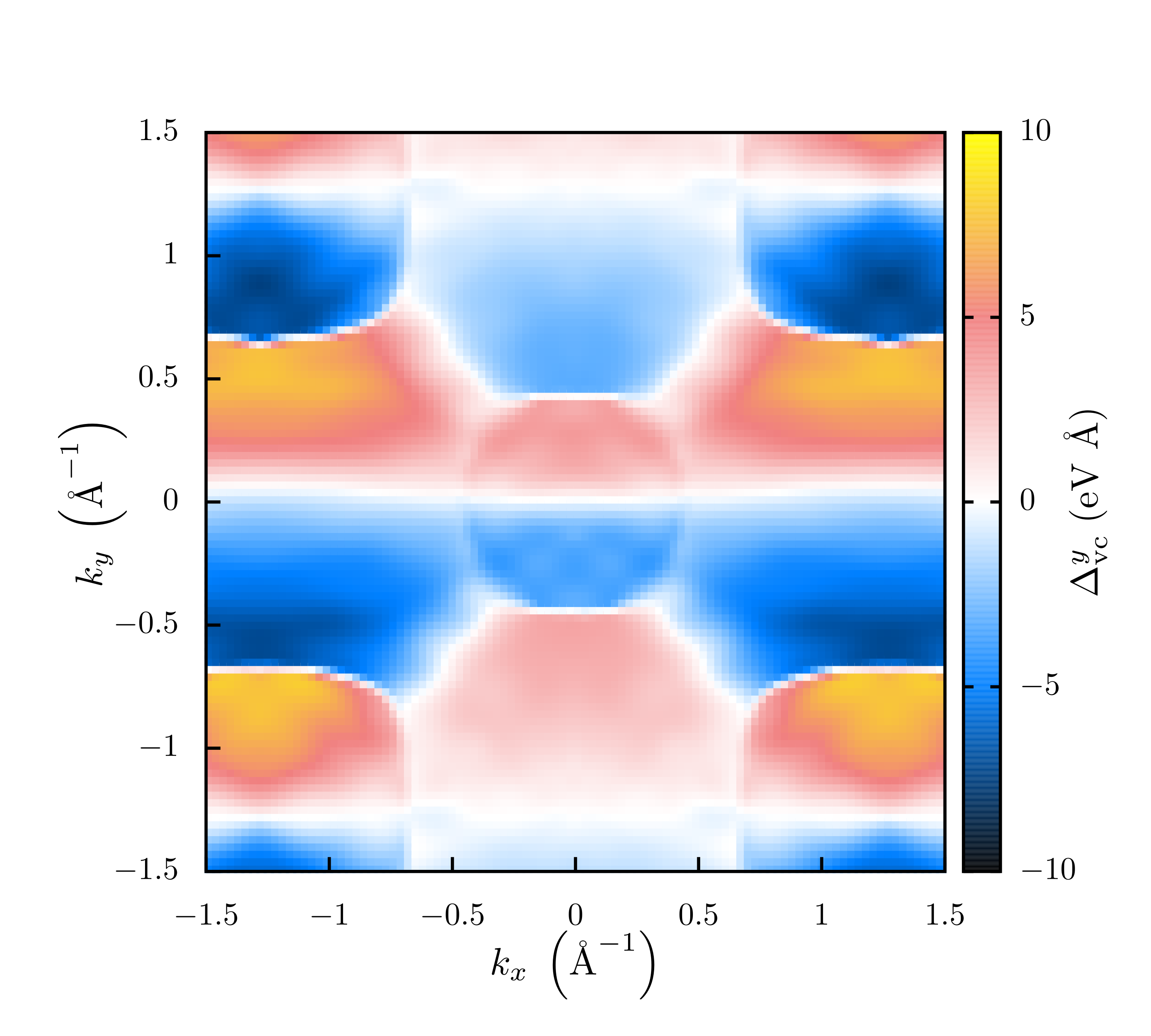}
\caption{Map of the velocity term ${\Delta^y_{\text{c}\text{v}}}$ at $k_{z}=0$ entering the injection-current matrix elements of Eq. (\ref{eq:inj_photocurrent}) for MF-I ($\bm{M}\parallel \hat x$).}
\label{fig:2d_velocity} 
\end{figure}

In Fig.~\ref{fig:2d_velocity} we show the calculated map of the velocity term ${\Delta^y_{\text{c}\text{v}}}$ at $k_{z}=0$ entering the injection-current matrix elements of Eq. (\ref{eq:inj_photocurrent}). The figure shows an asymmetric distribution around the $k_{y}=0$ line, which is the necessary condition for obtaining a nonzero injection current upon integration over the BZ. However, it is evident that the velocity term alone does not account for the hot spots around the $\mathrm{X}$ and $\overline{\mathrm{X}}$ points revealed by Fig.~3c in the Letter. These features must therefore originate from the dipole term of Eq. (\ref{eq:inj_photocurrent}).

\subsubsection*{MF-II structure}

For MF-II $(\bm{M}\parallel \hat z)$ there are four non-zero components of the symmetric components of the injection photoconductivity tensor, but only one of them is independent: $\eta_\mathrm{S}^{xyz}=\eta_\mathrm{S}^{xzy}=-\eta_\mathrm{S}^{yxz}=-\eta_\mathrm{S}^{yzx}$. The spectrum of $\eta_\mathrm{S}^{xyz}$ is presented in Fig. 4 of the Letter. It displays a negative peak at the onset of the visible energy range with maximum absolute value of $\sim2.5\times10^{8}$ A/(V$^2$s).

\subsection{Results for the antisymmetric part of the injection photoconductivity tensor}\label{sec.result_Ainjection}

\begin{figure}[h!]
\centering
\includegraphics[width=0.75\linewidth]{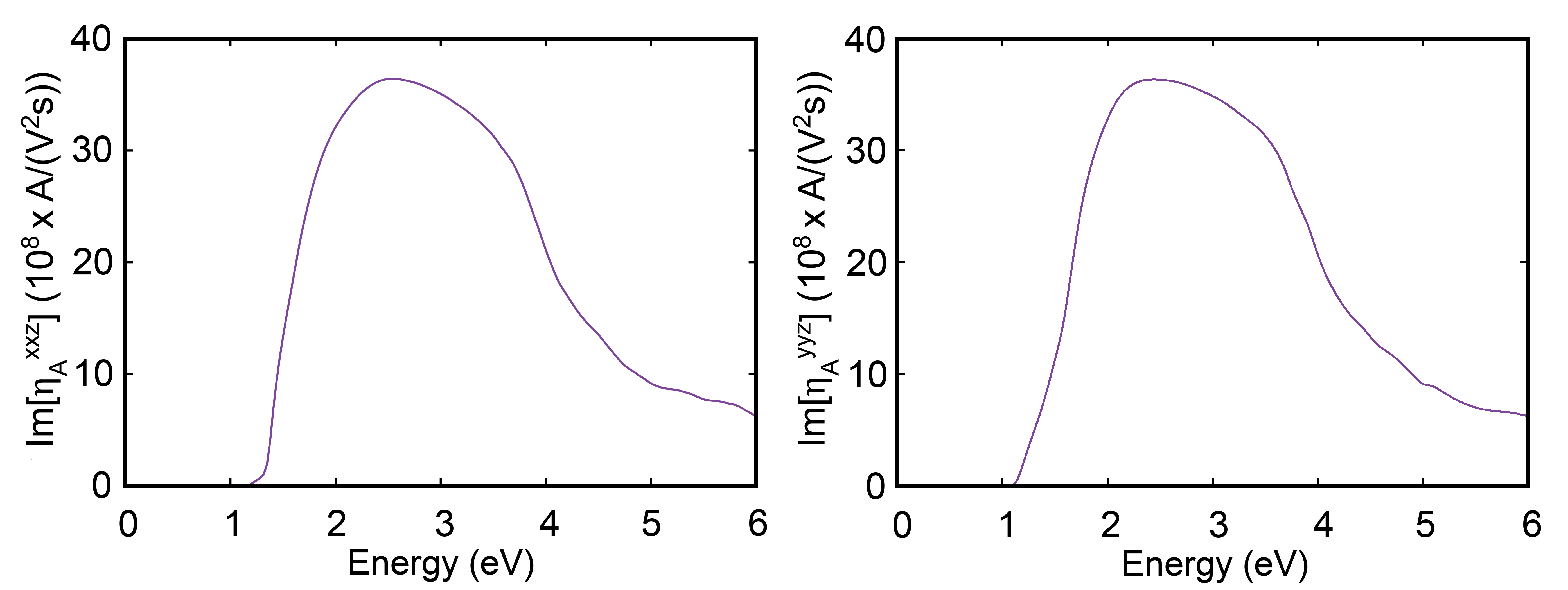}
\caption{Spectra of the non-zero antisymmetric components, $\eta_\mathrm{A}^{xxz}$ and $\eta_\mathrm{A}^{yyz}$, of the injection photoconductivity tensor of MF-I ($\bm{M}\parallel x$)}
\label{fig:CircularInjection} 
\end{figure}

 MF-I $(\bm{M}\parallel \hat x)$ has two non-zero and independent components of the antisymmetric part of the injection photoconductivity tensor, namely $\eta_\mathrm{A}^{xxz}$ and $\eta_\mathrm{A}^{yyz}$, with their spectra presented in Fig. \ref{fig:CircularInjection}.
 They are nearly identical and positive throughout the whole visible energy range, exhibiting a very broad maximum of $\sim 38$ $\times10^{8}$ A$/($V$^2$s) around $2.2$ eV. Notably, this value of the antisymmetric part of the injection photoconductivity is much larger than that recently reported for the Weyl semi-metal TaIrTe$_4$ \cite{Uriona2023_si}, which is considered to have a giant BPE \cite{Ma2019_si}. As such, according to our calculations, EuO in its MF-I state exhibits an extraordinarily large injection current for both linearly and circularly polarized light.
 
\begin{figure}[h!]
\centering
\includegraphics[width=0.4\linewidth]{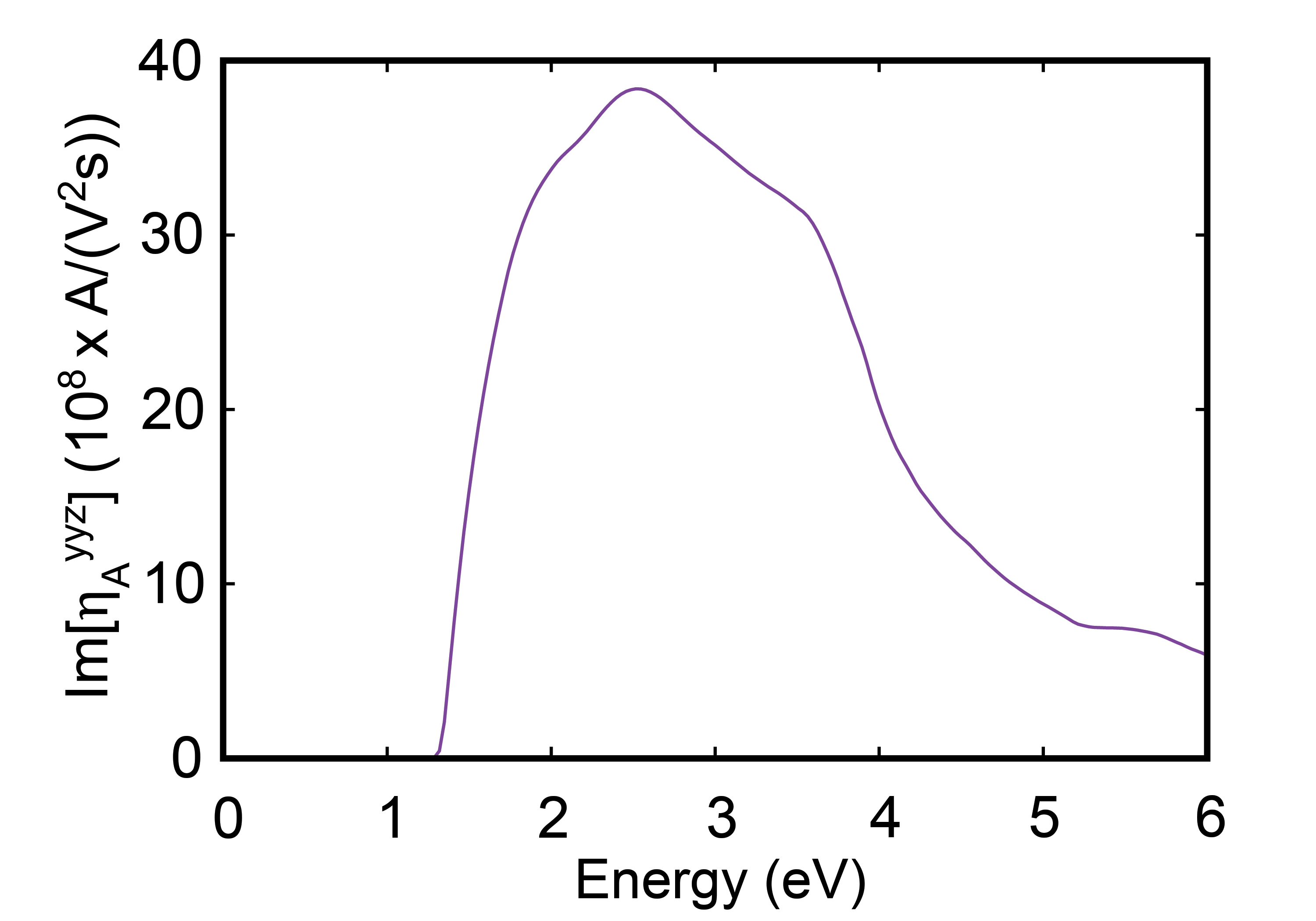}
\caption{Spectra of the non-zero antisymmetric component, $\eta_\mathrm{A}^{yyz}$, of the injection photoconductivity tensor of MF-II $(\mathbf{M}\parallel \hat z)$}
\label{fig:CircularInjectionMF2} 
\end{figure}

MF-II $(\bm{M}\parallel \hat z)$ exhibits the same two non-zero components of the circular injection photoconductivity tensor as MF-I. However, in MF-II, these components are not independent, but $\eta_\mathrm{A}^{xxz}=\eta_\mathrm{A}^{yyz}$. The corresponding spectrum,
shown in Fig. \ref{fig:CircularInjectionMF2}, resembles that of the non-zero components of MF-I, featuring a similar broad and large peak with a magnitude of $\sim 38$ $\times10^{8}$ A$/($V$^2$s) around $2.2$ eV. Thus, the circular injection photoconductivity appears to be nearly independent of the magnetic configuration.

\section*{Supplementary References}

\end{document}